# Exploring Thermal Transport in Electrochemical Energy Storage Systems Utilizing Two-Dimensional Materials: Prospects and Hurdles

Dibakar Datta, Eon Soo Lee

Department of Mechanical and Industrial Engineering
New Jersey Institute of Technology, Newark, NJ 07103, USA

Email: dibakar.datta@njit.edu, eonsoo.lee@njit.edu

**Abstract**

Two-dimensional (2D) materials (*e.g.,* graphene, transition metal dichalcogenides) and their heterostructures have enormous applications in Electrochemical Energy Storage Systems (EESS) such as batteries. A comprehensive and solid understanding of these materials' thermal transport and mechanism is essential for the practical design of EESS. Several advanced experimental techniques have been developed to measure the intrinsic thermal conductivity of materials. However, experiments have challenges in providing improved control and characterization of complex structures, especially for low-dimensional materials. Theoretical and simulation tools such as first-principles calculations, Boltzmann transport equations (BTE), molecular dynamics (MD) simulations, lattice dynamics (LD) simulation, and non-equilibrium Green's function (NEGF) provide reliable predictions of thermal conductivity and physical insights to understand the underlying thermal transport mechanism in materials. However, doing these calculations require high computational resources. The development of new materials synthesis technology and fast-growing demand for rapid and accurate prediction of physical properties require novel computational approaches. The machine learning (ML) method provides a promising solution to address such needs. This review details the recent development in atomistic/molecular studies and ML of thermal transport in EESS. The paper also addresses the latest significant experimental advances. However, designing the best low-dimensional materials-based heterostructures is like a multivariate optimization problem. For example, a particular heterostructure may be suitable for thermal transport but can have lower mechanical strength/stability. For bi/multilayer structures, the interlayer distance may influence the thermal transport properties and interlayer strength. Therefore, the review's last part addresses the future research direction in low-dimensional materials-based heterostructure design for thermal transport in EESS.

## 1. INTRODUCTION

In modern society, there is an urgent need for high-performance and cost-effective solutions to facilitate the widespread applications of rechargeable batteries in portable electronics, grid storage, renewable energy storage, and electric vehicles[1] . Batteries have become ubiquitous in our everyday lives. Most rechargeable battery products are built upon Lithium-Ion Batteries (LIBs), which were honored with the 2019 Nobel Prize in Chemistry[2]. However, the primary drawback of lithium-ion technology is its cost. Lithium, a finite resource on our planet, has experienced a reported price surge of approximately 738% since January 2021[3]. As an alternative to lithium, earth-abundant and cheaper metals such as aluminum (Al), calcium (Ca), magnesium (Mg), and Zinc (Zn) have been actively researched in battery systems[3-7]. These metals are multivalent (carrying two or more ionic charges). Hence, one ion insertion will deliver two or more electrons per ion during battery operation. Therefore, multivalent ions offer the promise of improved performance and cost efficiency. In summary, mono- and multivalent ions are being extensively investigated for sustainable energy storage.

The existing intercalation hosts present several challenges. For example, conventional graphite anode for lithium-ion batteries provides only a gravimetric capacity of 372 mAh/g[8]. On the other hand, the silicon anode for lithium-ion batteries can achieve a high capacity of 4000 mAh/g[9]. However, the major obstacle in utilizing silicon anodes for battery operations is the significant volume expansion of 300% during lithiation. In the case of beyond lithium batteries, higher charge density multivalent ions exhibit strong interactions with ions in the intercalation host, resulting in a relatively high migration barrier and slower diffusion kinetics. The multivalent ions insertion contributes to the host's mechanical degradation (stress-induced cracking), causing the material to lose its integrity. Consequently, mono- and multivalent ion-based batteries face significant challenges in identifying suitable hosts for energy storage.

In last few decades, the tremendous growth in the field of two-dimensional (2D) materials has shown great promise in energy storage because of their high specific surface areas offering electro-chemically active sites for ion storage and open 2D channels for fast ion transport[10, 11]. 2D materials bridge the gap between one-dimensional (1D) and three-dimensional (3D) bulk materials, raising new fundamental challenges associated with low-dimensional materials and opening a host of new applications. The intrinsic mechanical flexibility and common van der Waals (vdW) bonding in 2D materials make them suitable for seamless integration with existing active materials, thereby improving battery performance. Because of wide-ranging properties, 2D materials have potential applications in various aspects of batteries[12], including anodes, cathodes, conductive additives, electrode-electrolyte interfaces, separators, and electrolytes. Numerous studies have been conducted to explore the utilization of 2D materials and their heterostructures for energy storage[13].

A comprehensive knowledge of the thermal transport properties of materials and fundamental mechanisms[14-18] is crucial for a wide range of applications. These include energy storage[19, 20] , high-power density electronic devices[21, 22] , aerospace[23] , thermoelectrics[24, 25] , interfacial thermal management[26, 27] , thermal metamaterials[28] , and phonon engineering[29-31]. In the context of thermal conductivity, materials with low thermal conductivity can be utilized for thermal barrier coatings[32, 33] and thermoelectric power generators[15, 34] . On the other hand, materials with high thermal conductivity are suitable for efficient heat dissipation[35, 36] . Therefore, accurate characterization of materials' thermal conductivity is of utmost importance. One of the major reasons for battery failure is thermal runway. This phenomenon refers to a chain reaction within a battery cell that becomes challenging to halt once it initiates. It occurs when the temperature inside a battery reaches the point that triggers a chemical reaction within the battery. Thermal runway can lead to extremely high temperatures, violent cell venting, smoke, and eventually fire. Therefore, it is vital to study the thermal transport properties of battery electrode materials to effectively dissipate the heat generated during battery operation. This understanding can help prevent thermal runway and ensure the safe and efficient performance of batteries.

Several advanced experimental techniques[37-39] have been developed to measure the intrinsic thermal conductivity of materials. However, challenges still exist in these experiments, particularly when it comes to achieving improved control and characterization of complex structures, especially for two-dimensional materials. On the other hand, theoretical and simulation tools[40-42] have also proven to be valuable in predicting thermal conductivity and providing a deeper understanding of thermal transport mechanisms in materials. These tools include first-principles calculations, Boltzmann transport equations (BTE), molecular dynamics (MD) simulations, lattice dynamics (LD) simulation, and non-equilibrium Green's function (NEGF). These theoretical and simulation tools complement experimental techniques and contribute to our understanding of thermal transport in materials, including complex structures like 2D materials. They enable reliable predictions and provide physical insights that aid in the development and optimization of materials for various applications.

Conventional methods for calculating thermal transport properties can be time-consuming and present various challenges. Therefore, researchers are exploring the applications of machine learning (ML) models to predict thermal properties and overcome these limitations. Wu et al. [43] utilized different ML models to predict the interfacial thermal resistance and discovered that the Bi/Si material systems exhibited a high interfacial thermal resistance. This approach of combining ML models with traditional transport calculation methods represents a promising avenue for predicting the thermal conductivity of materials. In another study, Gu et al.[44] employed spectral neighbor analysis to investigate the lattice dynamics and thermal transport of single-layer $MoS_{2(1-x)}Se_{2x}$ alloys. The ML approaches offer efficient alternatives to conventional methods and open possibilities for exploring and optimizing thermal transport in various materials and systems.

Conventional bulk materials-based batteries face practical issues in meeting the ever-increasing energy demand. Thermal transport in 2D materials, such as graphene, has been studied extensively in the last decade. Newly developed 2D materials like TMDs (*e.g.,* $MoS_2$, $MoSe_2$), MXenes (*e.g.,* $Ti_3C_2$) are gaining attention for their thermal transport properties. Compared to most 3D materials, the low-dimensional structure and strong C-C covalent bonds in graphene result in distinct phonon dispersion and phonon-phonon scattering mechanisms, leading to high thermal conductivity ($\kappa$)[45, 46]. The measured room temperature $\kappa$ of mono-layer graphene is remarkably high[37], indicating an outstanding heat dissipation capability. However, measurement of suspended mono-layer graphene's $\kappa$ span a wide range[37, 47, 48, 49] of 1500-5400 W/mK. The $\kappa$ of mono-layer graphene supported by silica and copper substrates was found to be approximately 600 W/mK[50] and 50-900 W/mK, respectively. Different factors, such as defect concentrations, phonon-substrate scattering, and others, have been suggested to cause the significant variations in the measured $\kappa$.

Computational methods are highly beneficial in explaining numerous experimental observations. For instance, following pioneering experimental findings[47, 50, 51], theoretical calculations have been employed to reassess the layer-number dependence of graphene's thermal conductivity ($\kappa$)[52], investigate thermal rectification in mass-graded carbon nanotubes[53], and examine the substrate's impact[54] on graphene's $\kappa$. These computational studies have provided rigorous analyses of thermal transport in these structures and yielded mode-resolved phonon properties. A recent experiment[55] has further validated the width-dependent $\kappa$ of mono-layer graphene, as previously predicted by numerical studies[56-58].

Although there have been numerous studies on thermal transport in 2D materials, to the best of our knowledge, none of these studies have specifically focused on their utilization in energy storage and conversion systems such as batteries. Therefore, the findings from existing studies on 2D materials may not be directly applicable when considering their use in energy storage applications. For instance, when graphene is used as an anode in LIBs, it undergoes lithiation/delithiation process during charging and discharging. We propose that the thermal conductivity of lithiated graphene will differ from that of non-lithiated graphene. Furthermore, the number of lithium ions intercalated within the graphene structure may also influence its thermal transport characteristics. Similarly, the thermal conductivity of different 2D heterostructures is expected to vary at different charging and discharging states. The thermal conductivity of 2D materials may also differ when they encounter electrolytes or are integrated with other battery components. In summary, while there have been extensive studies on the thermal transport properties of 2D materials, their specific behavior, and characteristics in the context of energy storage systems such as batteries are yet to be fully explored. Therefore, further research is needed to understand and optimize the thermal properties of 2D materials in the context of their application in energy storage devices.

In this prospective review, we aim to address the challenging issue of understanding the thermal transport properties of 2D materials and their heterostructures when applied in energy storage and conversion systems, specifically batteries. The structure of the paper is organized as follows: **Section 2** discusses the existing problems associated with battery architecture and different failure modes that are commonly observed. **Section 3** highlights how 2D materials and their heterostructures can be implemented to overcome some of the common failure patterns observed in batteries. This section focuses on the potential benefits and advantages of using 2D materials for improving battery performance and reliability. **Section 4** discusses various methods of calculating thermal properties. We specifically explore different computational and theoretical approaches that can be employed to study the thermal transport characteristics of 2D materials in battery systems. **Section 5** provides a summary of the existing studies on the thermal properties of 2D materials. We review and consolidate the findings from various research efforts that have investigated the thermal conductivity and related properties of 2D materials in different configurations. **Section 6** explores more detailed experimental study of thermal transport in 2D materials. **Section 7** discusses the future directions and challenges in studying the thermal properties of 2D materials when used in energy storage such as batteries. We highlight the areas that require further research and propose potential strategies to address these challenges. **Finally**, paper concludes summarizing the novel key take-home messages. We believe that this prospective review will stimulate further theoretical and experimental work within the nanoscale heat transfer community, specifically in the context of 2D materials applied in energy storage systems.

## 2. PROBLEMS WITH THE EXISITING ENERGY STORAGE SYSTEMS

### 2.1 Traditional battery architecture

Figure 1 illustrates the traditional Lithium Ion Battery (LIB) architecture[12]. It should be noted that batteries based on other ions exhibit a similar structure. In this review, we will primarily focus on LIBs as a model system, but the discussions can be applied to other ion-based batteries as well. LIBs consist of two electrodes: a positive electrode and a negative electrode, along with an electrolyte. The electrolyte can either be organic, such as ethylene carbonate (EC), or aqueous based, such as Lithium bis(trifluoromethanesulfonyl)imide (LiTFSI). During the discharging process, lithium ions (cations) that are stored in the negative electrode (typically graphite) diffuse through the electrolyte and insert into the positive electrode material, such as $LiCoO_2$. Simultaneously, external electrons move towards the positive electrode. while external electrons move inward to the positive electrode. In the charging process, the movement of ions and electrons is reversed. To prevent electrical shorting between the electrodes while facilitating sufficient diffusion of lithium-ions, an ionically transparent and electrically insulating separator is sandwiched between the positive and negative electrodes. Traditionally, polymers have been used as binder in batteries. Additionally, materials like copper, nickel, etc. are commonly employed as current collectors. It is important to note that this review will consider the aforementioned LIB architecture, but the principles discussed can be extended to batteries based on different ions.

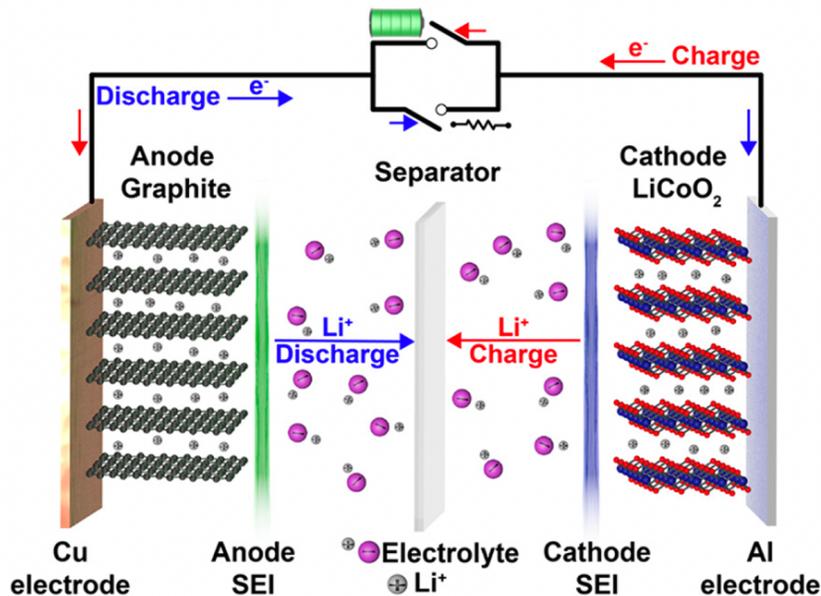

**Figure 1: Traditional battery configuration**. Re-printed with permission from ref[12].

## 2.2 Typical failure modes of batteries

There are many battery failure mechanisms, which needs a dedicated review. The two most significant chemo-mechanical failure modes observed in batteries (Figure 2a) are as follows:

**(i) Interface failure leading to electrical isolation of active particles**[59, 60] : The active particles, such as Silicon (Si) are in contact with metal current collectors (Figure 2b1) like nickel (Ni), which ensures a uniform electron distribution[61]. However, during the intercalation process, the active particles undergo significant volume expansion[62]. For instance, Si can experience a volume expansion of up to 300% upon lithiation[9, 63]. The metal current collectors, acting as non-slippery surfaces, are unable to accommodate this volume expansion/contraction of the active particles, which results in the generation of excessive interfacial stress (Figure 2b1) leading to fracture of active particles, ultimately causing battery failure[64].

Another critical interface within the battery system is between the polymer binder and active particles[60] (Figure 2b2). The polymer binder network keeps active particles together and provides a pathway for electron transport throughput the electrode[65]. However, volume expansion/contraction of active particles during charging/discharging generates excessive interfacial stress and fracture, leading to the electrical isolation of active particles[60] (Figure 2b2). When the active particles become electrically isolated, they are no longer able to contribute to the battery's overall electrochemical reactions, which leads to a decline in battery performance.

Addressing these issues related to interfacial stress and fracture of active particles is crucial for improving the performance and reliability of batteries. Strategies aimed at mitigating these challenges are actively sought after to enhance the overall functionality and lifespan of battery systems.

**(ii) Failure of active particles[66-69]**: In addition to interface failure, the fracture of active particles is indeed a significant factor contributing to battery failure[70]. For instance, the volume expansion and contraction experienced by Si anode particles during lithiation and delithiation cycles can lead to their fracture[71], which ultimately results in battery failure[72]. We urgently need strategy to address these practical problems.

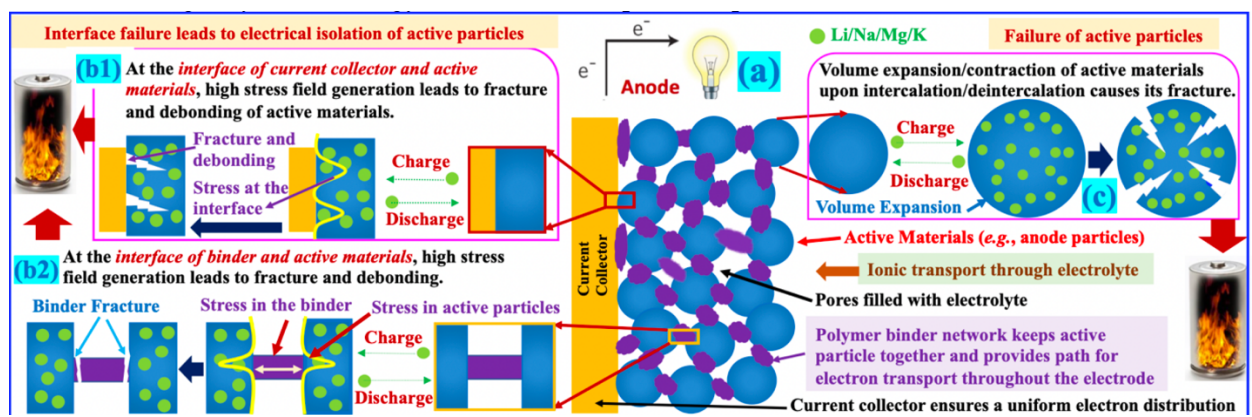

**Figure 2: Current failure modes in battery: [a]** Schematic of anode, **[b]** Different interface failure modes, **[c]** Failure of active particles. Re-printed with permission from ref[73].

## 3. 2D MATERIALS FOR ENERGY STORAGE

### 3.1 Overview of 2D materials

Figure 3 illustrates various 2D materials and their heterostructures (2D + $n$D, where n = 0, 1, 2, 3). These materials have wide-ranging applications across multiple fields (Figure 3b)[74]. The field of 2D materials was initiated with the isolation of monolayer graphene flakes from bulk graphite through mechanical exfoliation[75, 76]. Since then, numerous 2D materials have been discovered, including transition metal dichalcogenides (TMDs) such as $MoS_2$[77], hexagonal boron nitride (h-BN)[78], black phosphorous (BP) (or phosphorene)[79], and MXenes (*e.g.,* $Ti_3C_2$)[80]. The 2D materials family offers a diverse range of physical properties. Graphene, for instance, exhibits excellent conductivity, while materials like $MoS_2$ demonstrate semiconducting behavior, and h-BN exhibits insulating properties[81]. By 2020, more than a thousand 2D materials have been identified[82], opening possibilities for novel physics and applications in various fields, including energy storage.

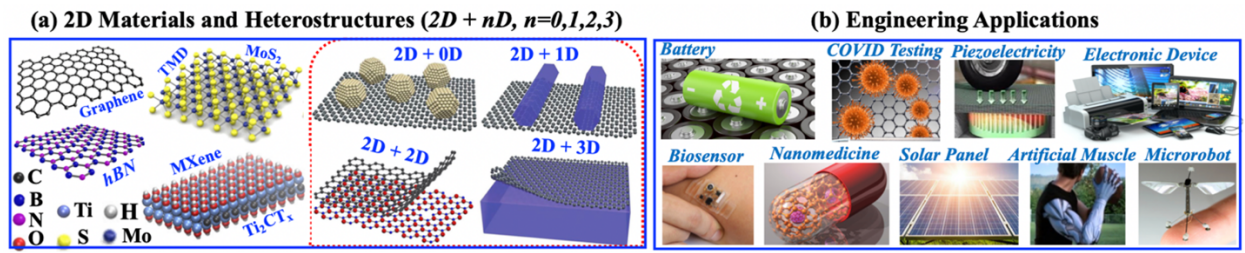

**Figure 3: 2D materials and their applications.**

## 3.2 Overview on how 2D materials can be useful for battery applications

Figure 4 provides a comparison between 3D and 2D materials-based electrodes in terms of their performance in batteries[13].

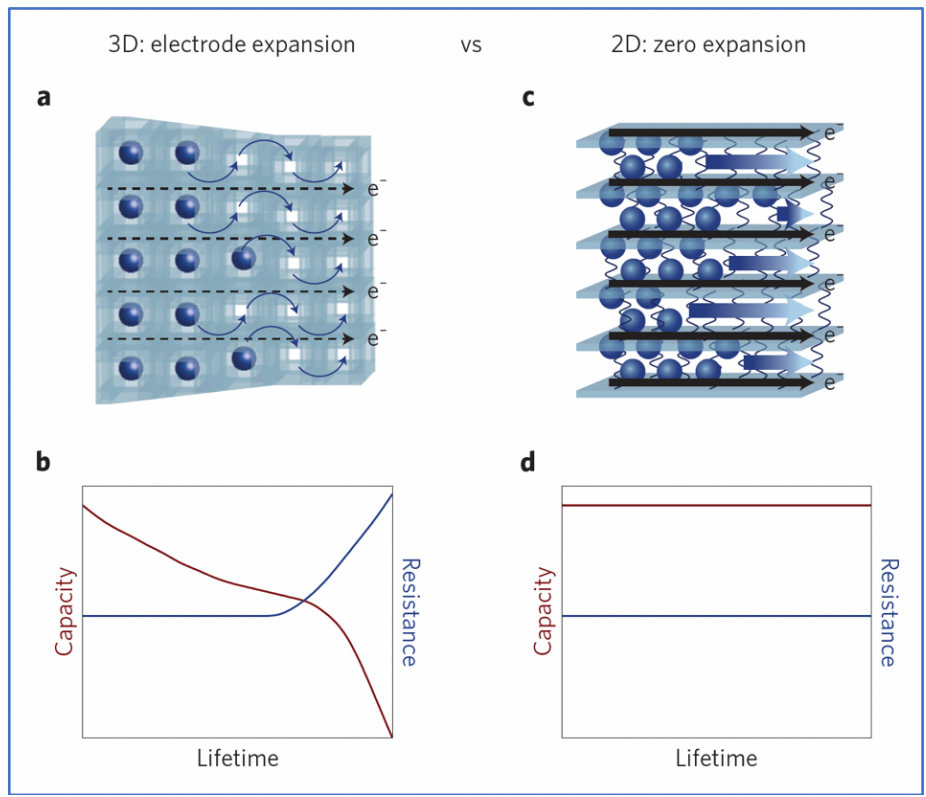

**Figure 4: Overcoming limitations of current batteries by using 2D materials. [a]** Schematic of a 3D intercalation electrode, and **[b]** increase in resistance and decrease in capacity because of mechanical and/or structural degradation of electrode materials. **[c]** A zero-expansion 2D electrode with improved kinetics of electrochemical processes due to easy transport of electrons and ions, and **[d]** stable electrochemical performance and extended lifetime due to the close-to-zero volume change of pillared 2D electrodes. Re-printed with permission from ref[13].

In a 3D electrode architecture, battery capacity tends to fade over time, and the resistance of the electrode increases. This can lead to a reduction in the overall performance and lifetime of the battery. On the other hand, in a 2D scenario, when proper design strategies are implemented, it is possible to retain the capacity of the battery over time. This implies that 2D materials have the potential for extensive applications in battery design. The comparison depicted in Figure 4 highlights the advantages offered by 2D materials in terms of their ability to maintain capacity and performance over the lifetime of a battery. These findings suggest that incorporating 2D materials into battery electrode designs can lead to enhanced battery performance and longevity[83-88].

### 3.2.1 Prevention of interface failure:

**2D materials as binder:** One potential solution to address the challenges associated with the polymer binder in battery electrodes is to replace it with 2D MXenes (Figure 3a1)[89]. We hypothesize that 2D materials, such as MXenes, can act as a 'slippery' surface, thereby minimizing interfacial stress. By replacing the traditional polymer binder with 2D MXenes, it is possible to reduce the friction and stress at the interface between the active particles and the binder. The unique properties of MXenes, including their high mechanical strength and flexibility, make them promising candidates for serving as a binder material. These 2D materials can offer enhanced adhesion and mechanical stability, reducing the likelihood stress and failure[64].

**2D materials as current collector:** To address the interface failure between the current collector and active particles (Figure 2b1), we propose two potential options:

*(i) Addition of a 2D material, such as graphene, as a 'coating' on the current collector to create a 'slippery' interface (Figure 5a2)[64, 90]:* By applying a thin layer of graphene or another suitable 2D material onto the surface of the current collector, it can act as a lubricating layer, reducing interfacial stress and friction between the current collector and the active particles. This approach is to improve the mechanical compatibility and alleviate stress concentration at the interface.

*(ii) Complete replacement of the current collector with a 2D materials, such as MXenes (Figure 5a3)[91, 92]:* In this option, the traditional current collector material, such as copper or nickel, would be entirely substituted with a 2D materials like MXenes. MXenes, known for their excellent conductivity and mechanical properties, can serve as both the current collector and a structural support for the active particles. This replacement eliminates the interface between the current collector and active particles, potentially reducing interfacial stress and improving overall electrode performance.

Both options aim to mitigate the interface failure between the current collector and the active particles by leveraging the unique properties of 2D materials. Whether by coating the current collector or completely replacing it with a 2D material, these approaches can help alleviate stress concentration and enhance the mechanical stability of the electrode interface, ultimately improving battery performance and reliability.

### 3.2.2 Prevention of active materials failure:

To solve the problem of active materials failure (Figure 2c), we propose two approaches:

*(i) Replace 3D active materials (e.g., Si) with 2D materials*[8, 10, 93] *and their heterostructures (e.g., MXenes + MoS$_2$, Graphene + MoS$_2$) (Figure 5b1)*[13]: By substituting 3D active materials with suitable 2D materials, we can potentially overcome the limitations and failure modes associated with 3D materials. 2D materials offer unique properties that can improve the stability, mechanical flexibility, and electrochemical performance of battery electrodes. Additionally, the integration of heterostructures, combining different 2D materials, can further enhance the functionality and properties of the active materials.

*(ii) Integrate 3D active materials with 2D materials (Figure 5b2)*[94]: Instead of completely replacing the 3D active materials, this approach involves combining them with 2D materials. By integrating 3D active materials with appropriate 2D materials, we can take advantage of the strengths of both material types. The 2D materials can act as a protective layer, coating, or interface modifier, enhancing the stability, mechanical properties, and electrochemical performance of the 3D active materials.

In all the proposed approaches, interfacial mechanics plays a critical role in electrochemical performance, *i.e.*, energy and power density, volumetric capacity, etc. The mechanical integrity of these interfaces (*e.g.*, 2D + 2D, 2D + 3D) dictates the long-term performances of energy storage systems[95, 96].

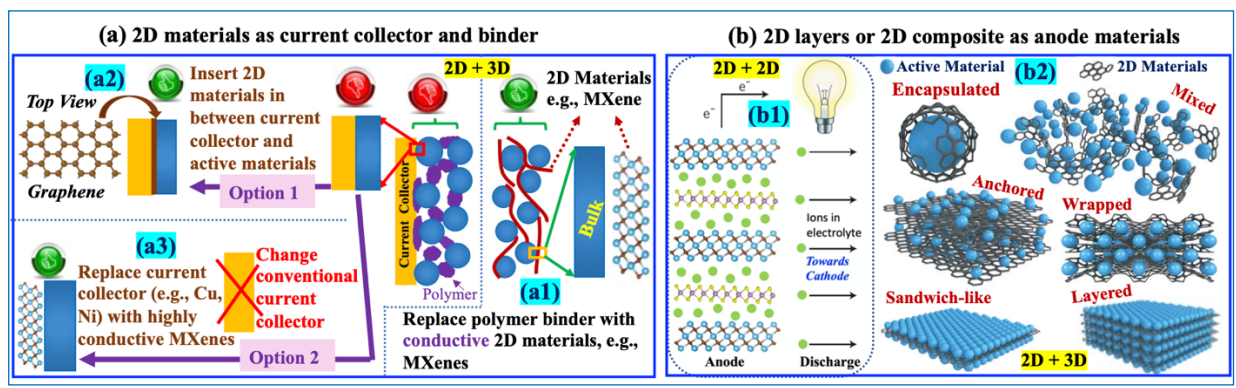

**Figure 5: 2D materials for anode, current collector, binder: [a]** 2D materials as current collector and binder, **[b]** 2D layers and 2D composite as anode materials. Re-printed with permission from ref[73]

### 3.3 2D materials as electrode

Figure 6 shows the usage of 2D graphene for applications in energy storage[8]. The mechanism of defect-induced storage of metallic lithium within a porous structural network (PGN) has been extended to synthesize a high-performance electrode material. Figure 6a shows the PGN electrode. The experimental voltage-capacity (Figure 6b) shows that PGN can achieve up to 1000 mAh/g, almost 3 times higher than conventional graphite (372 mAh/g). The experimental findings were verified by theoretical research. Figure 6c shows the DFT calculations on graphene: pristine (c1), divacancy defect (c2), stone-wales defect (c3). The pristine graphene is unsuitable for adatom adsorption. However, defects in graphene facilitate adatom adsorption. Figure 6d shows that the capacity of defective graphene-based electrode can be 3-4 times higher than conventional graphite.

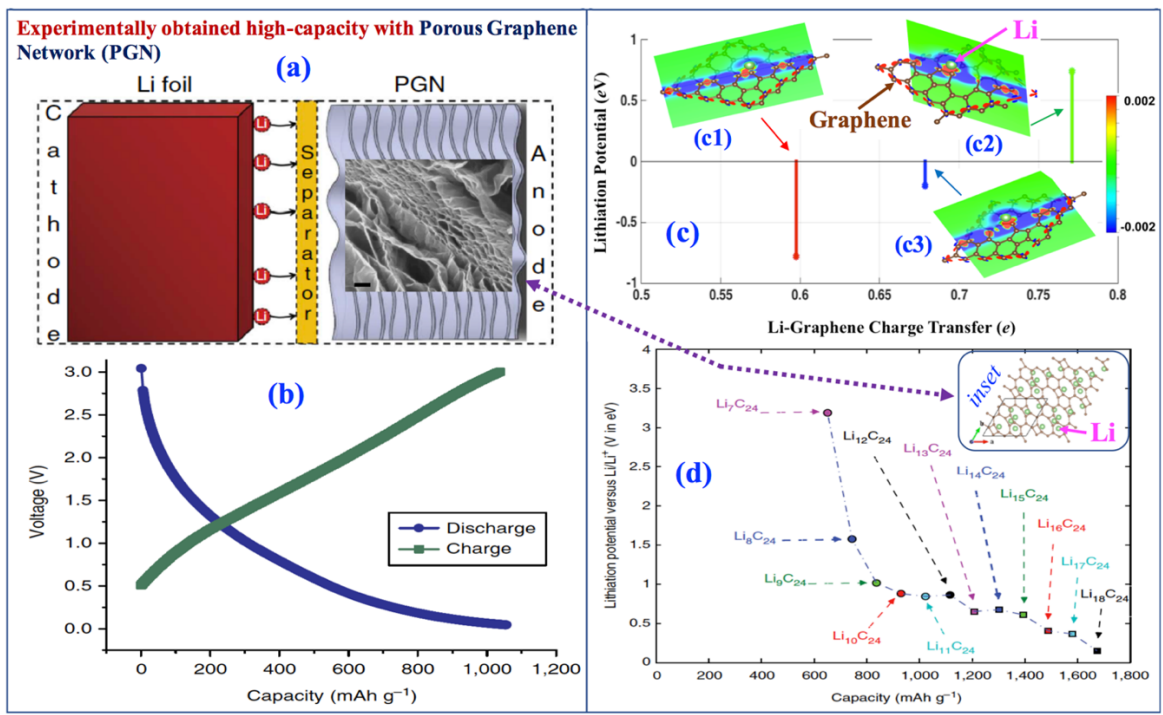

**Figure 6: Defective graphene as high-capacity anode materials.** [**a-b**] Experimental work showing the porous graphene network (PGN) for high-capacity anode. [**c**] DFT study of Li adsorption on graphene – (c1) pristine, (c2) stone-wale, (c3) divacancy. [**d**] DFT study of capacity-potential curve for defective graphene. Re-printed with permission from ref[8].

### 3.4 2D materials as vdW slippery interface over current collector

**Interface adhesion:**

Figure 7 shows the interfacial adhesion (work of separation, $W_{sep}$) of several amorphous silicon ($a$-Si) and substrate combinations[68]. The $W_{sep}$ for $a$-Si/graphene system is around one-fifth of that for the Cu and Ni substrates and less than half of that for the $a$-Si/$a$-Si substrate. This indicates that compared to traditional metal current collectors (*e.g.,* Cu, Ni), graphene offers a much slippery interface with the $a$-Si film.

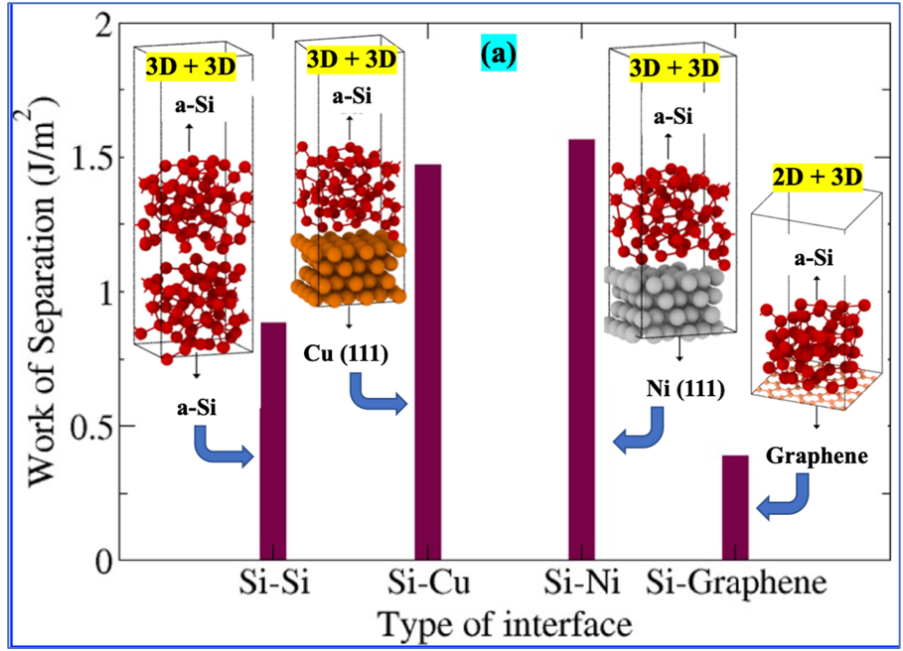

**Figure 7: Work of Separation (J/m²) for 3D-3D and 3D-2D interfaces.** Reprinted with permission from ref[64].

**Interface Stress:**

The large volume expansion of silicon leads to stress build-up at the interface between Si film and current collector, leading to delamination of Si from the surface of the current collector. 2D materials graphene can be engineered as a vdW 'slippery' interface between the Si film and the current collector. This can be achieved by simply coating the current collector surface with graphene sheets. Figure 8 shows MD simulation[64] of lithiation and delithiation of *a*-Si for two cases: *a*-Si/*a*-Si (Figure 8a0) and *a*-Si/graphene (Figure 8b0). Figures 8a1 and 8b1 show lithiation and delithiation for these two cases. Figure 8a2 and 82 show the shear stress ($\tau_{xx}$) profile at lithiation (Li/Si = 3.5, black line) and delithiation (Li/Si = 0.3, orange line) for the system with nonslip (Figure 8a0) and slip (Figure 8b0) surface. For graphene slippery surface (Figure 8b0), lithiated *a*-Si does not stick to the surface (Figure 8b1). Hence the stress at the interface is reduced (Figure 8b2).

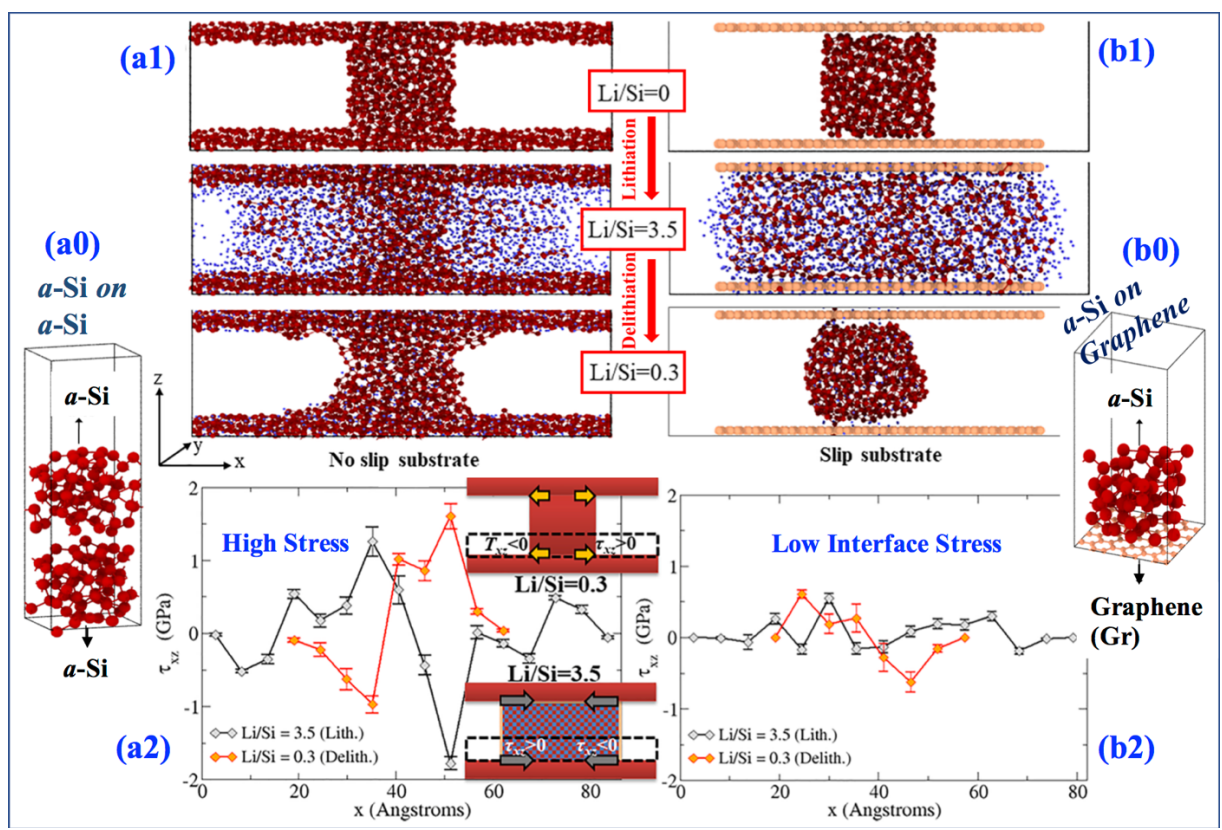

**Figure 8: [a0]** Silicon on silicon substrate. **[a1, b1]** a1 and b1 show snapshots of the simulation systems with a rigid nonslip substrate (a1) and a rigid slip substrate (b1) during lithiation and delithiation cycle. The snapshots are taken prior to lithiation, at a highly lithiated stage (Li/Si = 3.5) and a highly delithiated stage (Li/Si = 0.3), from top to bottom. The red atoms are silicon, while the blue atoms are lithium. **[a2, b2]** shear stress $\tau_{xz}$ profile at Li/Si = 3.5 (black lines) and Li/Si = 0.3 (orange lines) for the systems with nonslip (a2) and slip substrate (b2), respectively. $\tau_{xz}$ is averaged over the region within 1 nm above the bottom substrate (as indicated by the box) every 0.55 nm along the x-direction. The error bar shows the stress fluctuation over 10 independent stress measurements. Reprinted with permission from ref[64].

### 3.5 2D materials for other aspects in batteries

2D materials can be *implemented as conductive additives*. One of the most effective strategies has been to exploit the high aspect ratio of graphene to encapsulate active materials, resulting in improved electrochemical properties and stability[97-99]. 2D materials can be implemented to *control the electrode-electrolyte interface.* Ultrathin 2D materials stable interfacial layer can suppress dendritic growth and uncontrolled SEI (Solid Electrolyte Interphase) formation[100]. 2D materials can also be utilized as *separators and electrolytes*. The primary purpose of separators in LIBs is to prevent electrical contact between the cathode and anode without compromising lithium-ion diffusion. Therefore, mechanical robustness and ion diffusivity are important criteria for separator. 2D materials have shown to be promising candidates for this purpose[101].

### 3.6 2D materials heterostructures for energy storage

Figure 9 shows the 2D materials heterostructures for energy storage[13].

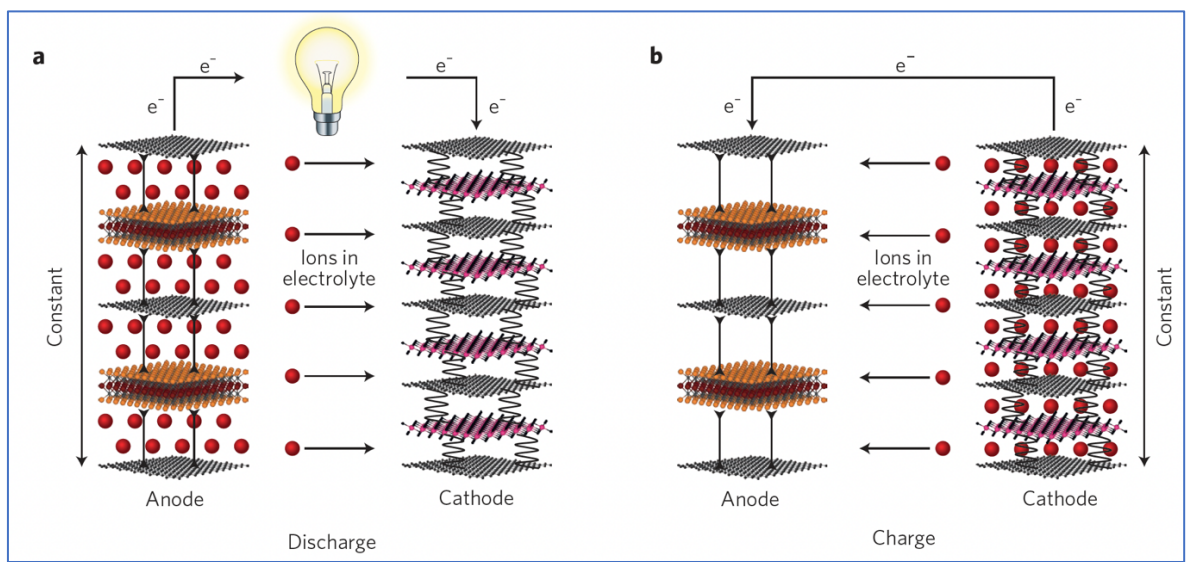

**Figure 9: Schematic illustration of the electrochemical cycling process in a battery with 2D heterostructured pillar electrodes: [a,b]** Battery discharge cycle (a) and charge cycle (b). Pillaring can be used as an effective strategy to conserve the interlayer distance in 2D heterostructures enabling zero-volume change in both electrodes during battery cycling, thus improving mechanical and electrochemical stability of the system leading to extended cycle life. The pillars are schematically represented by elements, shown in black, connecting layers in the 2D heterostructured anode and cathode. Reprinted with permission from ref[13].

Some key points regarding 2D heterostructures are:

1. *Advantages of Heterostructured Architectures:* Individual 2D materials have specific properties that may be advantageous for energy storage, but often they lack other desirable properties. Stacking different 2D materials in heterostructures allows for combining the advantages of different building blocks while mitigating their shortcoming.

2. *Specific Surface Area and Ion Transfer:* The large specific surface area of 2D materials in heterostructures enhances the contact area with the electrolyte, leading to improved kinetics of ion transfer. However, excessive consumption of electrolyte during the formation of a solid-electrolyte interface (SEI) layer can result in irreversible capacity loss in the first cycle.

3. *Control of Interactions and Assembly:* The strength of interaction and distance between the layers in heterostructures can be controlled through chemical modifications and assembly parameters. For example, weak interactions between graphene and MXene layers facilitate intercalation and faster ion diffusion at the interface compared to interfaces between layers of the same material.

4. *Influence on Electronic Properties:* The direct contact between layers in heterostructures can significantly affect electronic properties. Introducing stabilizing species, such as ions and molecules, at the heterointerface can enhance electrochemical stability and ion diffusion.

5. *Investigating Different Interface Architectures:* Both intimate (direct) contact between layers and through-interlayer-species (quasi-intimate) interfaces should be explored to determine the most efficient stacked electrode architecture in 2D heterostructures.

Overall, the combination of different 2D materials in heterostructured architectures offers the potential to optimize energy storage performance by leveraging the unique properties of induvial components while addressing their limitations. Exploring and understanding the interplay between interfacial properties, ion transport, and electronic properties in 2D heterostructures is essential for developing highly efficient energy storage systems.

## 4. COMPUTATIONAL METHODS FOR THERMAL TRANSPORT STUDIES

### 4.1 Brief Overview of Computational Methods for Thermal Transport Calculations

Several theoretical methods have been developed to study thermal transport in nanomaterials such as Boltzmann transport equation (BTE), Atomistic Green's functions (AGF), Molecular Dynamics (MD) simulations. The AGF and MD methods are suitable for describing phonon transmission/scattering across an interface in 2D heterostructures. These methods can capture the thermal transport behavior from all phonons modes as a whole or track the contribution from a single phonon mode based on its relaxation time, mean free path (MFP), and velocity. They are appropriate for different phonon transport regimes or aspects because they deal with phonons in different manners, such as wave versus particle nature, time versus frequency domain, different thermodynamic conditions, and different boundary conditions (BCs). Here, we briefly summarize different computational methods.

**Classical Molecular Dynamics Simulations**

MD simulations are indeed widely used to study thermal transport in 2D materials. MD simulations can account for various factors such as doping, defects, strain, and substrate effects on thermal transport. The accuracy of MD results relies on the quality of the interatomic potential, and significant efforts are being made to develop reliable MD potentials for accurate simulations[102]. Classical MD simulations model the movements of atoms based on Newton's second law of motion and interatomic potentials. MD simulations can directly capture phonon thermal transport and naturally incorporate atomic-level details, including defects, interfaces, strain, surface reconstruction, and other structural characteristics. This atomic-level resolution provides valuable insights into the mechanisms and behavior of thermal transport in 2D materials.

Several MD schemes have been employed to model heat transfer, including Nonequilibrium MD (NEMD)[103], Reverse NEMD (RNEMD)[104], Equilibrium MD (EMD)[103], Thermal relaxation[105], The wave-packet (WP) method[106], Phonon normal mode analysis (NMA)[107, 108]. The accuracy of MD is limited by the quality of empirical interatomic potentials (EIP). To overcome this limitation, first-principles MD, which relies on ab initio calculations and accurate interatomic potentials derived from quantum mechanics[109] has also been applied to thermal modeling in recent years[110].

The equilibrium MD (EMD) and nonequilibrium MD (NEMD) simulations are commonly used to determine the thermal properties of 2D materials. In NEMD simulations, a thermal gradient is imposed by connecting both sides of sample to hot and cold reservoirs. During the simulation, the system reaches a stationary state where the heat flux and temperature become constant. This allows for the calculation of thermal conductivity based on Fourier's law of heat conduction, which relates the heat flux ($J$) to the temperature gradient ($\nabla T$) through the material.

Based on Fourier's law, $\kappa$ can be computed as:

$$\kappa = \frac{J}{\nabla T A} \tag{1}$$

Where A is the cross-sectional area of the simulated cell. In NEMD simulations, to maintain a constant temperature gradient ($\nabla T$) or heat current ($J$) across the simulation cell, two thermostats are typically employed. These thermostats control the temperature at the boundaries of the system, ensuring the desired temperature gradient or heat current is maintained.

The NEMD method has been widely used not only to study bulk thermal properties but also to investigate interface-modulated phonon dynamics in various 2D materials systems. For example, NEMD simulations have been employed to explore the thermal transport characteristics of graphene/$MoS_2$[111] and graphene/phosphorene in-plane heterostructures[112]. These simulations provide insights into the thermal conductivity and phonon scattering mechanisms at the interface between 2D materials. Similarly, NEMD simulations have been applied to investigate thermal properties of graphene/h-BN[113], $MoS_2$/$WSe_2$[114], and phosphorene/graphene van der Waals (vdW) heterostructures[115]. By employing NEMD simulations, researchers can gain a deeper understanding of how interfaces between different 2D materials influence phonon dynamics and thermal transport.

**Green-Kubo Method (GK)**

The Green-Kubo method is commonly employed in Equilibrium Molecular Dynamics (EMD) simulations for the calculation of thermal conductivity. In EMD, the thermal conductivity is calculated using Green-Kubo formulation or the Einstein relation[116, 117]. This method utilizes the fluctuation-dissipation theorem[103, 118] to relate the heat current fluctuations in the system to the thermal conductivity. Specifically, for a system in the $x$-direction, $\kappa$ can be computed using the following equation –

$$\kappa_x = \frac{1}{k_B V T^2} \int_0^\infty \langle J_x(t) J_x(0) \rangle \, dt \tag{2}$$

Where V, $k_B$, t, and $J_x$ denote the volume of the simulation cell, Boltzmann constant, time, and heat current in the x direction. $\langle J_x(t) J_x(0) \rangle$ denotes the heat current autocorrelation function (HCACF). The Green-Kubo method provides a statistical approach to determine thermal conductivity in EMD simulations. By analyzing the fluctuations of the heat current, it offers a means to extract the macroscopic transport property of the system.

**Phonon Wave-Packet Method**

The Wave-Packet Propagation (WPP) method[106] is a computational approach used to capture the dynamic propagation and scattering of phonons by various features such as boundaries, interfaces, and defects. This method simulates the behavior of phonons by initializing atomic displacements from their equilibrium positions according to a formula that represents a wave packet centered at a specific phonon mode $(\kappa, \omega)$ and positioned at a specific location within the material. In the WPP method, the wave packet is then allowed to propagate through the material, following the group velocity of the corresponding phonon mode. As the wave packet encounters scattering centers such as boundaries, interfaces, or defects, its dynamics are affected, leading to scattering events. By computing the total energy of the wave packet before and after each scattering event, the transmission coefficient can be evaluated.

**Normal Mode Analysis (NMA)**

NMA (Normal Mode Analysis) is a commonly used method in EMD simulation, including GK-MD (Green-Kubo Molecular Dynamics). NMA can be utilized to compute $\kappa$ by incorporating the extracted phonon dispersion and spectral phonon relaxation time[119] $\tau$. The $\tau$ prediction in NMA can be accomplished in both the time domain and the frequency domain. In the time domain[107, 120], $\tau$ can be estimated by calculating the decay rate of the spectral energy, reflecting the relaxation of phonon modes. In the frequency domain[121], $\tau$ can be obtained by evaluating the linewidth of the spectral energy, which is often referred to as spectral energy density (SED) analysis.

By utilizing the predicted mode-wise $\tau$ and the group velocity ($v_g$) derived from the phonon dispersion, $\kappa$ can be evaluated under the Relaxation Time Approximation (RTA). The RTA assumes that phonon scattering processes are dominated by binary collisions with an average relaxation time. Therefore, the thermal conductivity can be expressed as:

$$\kappa_x = \sum_k \sum_v c(k,v) v_{g,x}^2(k,v) \tau(k,v) \qquad (3)$$

Here, the summations are performed over all phonon modes and branches. The subscript $x$ indicates the longitudinal direction. NMA provides a comprehensive approach to study thermal transport in materials by considering the dispersion and relaxation characteristics of phonons. It enables the estimation of thermal conductivity based on mode-wise relaxation times and group velocities, offering insights into the underlying mechanisms governing heat transfer in materials.

**Nonequilibrium Green's Function Method or Atomistic Green's Functions (AGF)**

The Atomistic Green's Function (AGF) method is a powerful computational approach that rigorously accounts for the wave nature of phonon on a discrete atomic lattice. Originally developed for quantum electron transport in nanostructures[122, 123], the AGF method was later extended to study phonon transport and thermal conductivity in nanostructures[124-126]. The AGF method is particularly well-suited for investigating thermal transport in low-dimensional heterostructures, such as Si/Ge[127], graphene/hBN[128], $MoS_2$/metal[129], interfaces, and others. It enables a detailed understanding of phonon propagation and scattering processes in these systems.

Unlike classical MD simulations that consider phonons as classical particles, the AGF method utilizes the principles of quantum mechanics and incorporates the Bose-Einstein statistics to treat phonon distributions. As a result, AGF method is valid even at sub-Debye temperature $\Theta_D$, where the quantum behavior of phonons becomes significant. In the AGF method, the transmission function ($\Xi(\omega)$) across the system is computed based on the Green's functions, which are constructed from the interatomic force constants of the system.

The heat flux and the thermal conductance can be computed using $\Xi(\omega)$ by the Landauer formula[130]:

$$J = \int \frac{\hbar\omega}{2\pi} \Xi(\omega)[n_1(\omega, T_1) - n_2(\omega, T_2)]d\omega \tag{4}$$

$$G = \left| \frac{1}{T_1 - T_2} \int \frac{\hbar\omega}{2\pi} \Xi(\omega)[n_1(\omega, T_1) - n_2(\omega, T_2)]d\omega \right| \tag{5}$$

The phonon occupation numbers of the two leads, denoted as $n_1$ and $n_2$, correspond to the thermal equilibrium states of the leads. These occupation numbers are associated with the temperatures $T_1$ and $T_2$ of the two leads, respectively. The AGF method considers the phonon transport across the system by considering the phonon distribution in the leads. This method can be referred to as "first principles" when the interatomic force constants used in the calculations are extracted from density functional theory (DFT).

**Boltzmann Transport Equation (BTE)**

The Boltzmann Transport Equation (BTE) is a powerful tool used to study the statistical behavior of the non-equilibrium thermodynamic system, including the transport of phonons in materials. In the particle-based picture perspective, the BTE describes the behavior of phonons, which follow the Bose-Einstein distribution function $n_\lambda^0$ at thermal equilibrium but deviate from it under the influence of a temperature gradient. The phonon BTE for a specific phonon mode $i$ under a temperature gradient $\vec{\nabla}T$ can be expressed using perturbation theory:

$$-\vec{v}_{g,i} \cdot \vec{\nabla}T \frac{\partial n_i}{\partial T} + \left(\frac{\partial n_i}{\partial t}\right)_{collision} = 0 \tag{6}$$

The equation describes the balance of phonon population ($n$, occupation number) in a material, considering both the diffusion drift (first term) and collision (second term, also known as scattering) of phonons. It represents the transport behavior of phonons in response to temperature gradients and scattering processes[131].

### 4.2 Brief Overview of Thermal Transport Prediction by Machine Learning

Theoretical calculations and simulations play a crucial role in studying the lattice thermal conductivity ($\kappa_p$) of materials. However, the increasing complexity of materials and the demand for high-accuracy predictions pose significant computational challenges. To address these challenges, researchers have turned to machine learning (ML) techniques, which offer a novel avenue to improve accuracy and reduce computational costs.

There are various ML algorithms that have been successfully employed for the prediction of physical properties of materials. Some popular algorithms include Support Vector Machines (SVM)[132], Artificial Neural Networks (ANNs)[133], Decision Trees[134], Bayesian Optimization[135], etc. The ML models are trained using a dataset that contains known input features and corresponding target values. The model's parameters are then obtained through training, and the performance is evaluated using validation data. Hyperparameters, which control the behavior of the ML models, can be optimized using the validation data to enhance their performance.

ML potentials have emerged as a powerful approach to establish the relationship between atomic structures and the potential energy surface (PES) of a system. ML algorithms can be employed to map the input features, typically atomic positions, or configurations, to the corresponding output, which is the potential energy of the system. Once ML potentials are established, they can be used to compute atomic forces and system total energies directly without the need for computationally expensive first principles calculation or empirical potentials. In the context of thermal transport calculations, ML potentials can be utilized as input information for various methods such as the Boltzmann transport equations and MD simulations. These calculations can provide insights into the phonon transport mechanisms involved in the system and contribute to a better understanding of thermal conductivity and related properties.

The Artificial Neural Networks (ANNs) are traditional ML algorithms that have been widely used in various fields. One significant application of ANNs is the development of neural-network potentials (NNPs) for MD simulations, as pioneered by Behler and Parrinello[136]. The success of ML models in predicting thermal conductivity relies on the selection of appropriate algorithm and descriptors. Different ML algorithms can yield varying performances depending on the dataset and the specific problem at hand. In the study by Wang et al.[137], four different ML models, including Support Vector Regression (SVR), Kernel Ridge Regression, eXtreme Gradient Boosting (XGBoost), and fully connected neural networks, were trained to predict the thermal conductivity of crystal materials using data from the Inorganic Crystal Structure Database (ICSD). Through their investigation, Wang et al. found that XGBoost outperformed the other ML models in predicting thermal conductivity for the search of materials with low thermal conductivity. XGBoost is a boosting algorithm that combines multiple weak prediction models (typically decision trees) to create a more accurate and robust model.

ML potentials received increasing attention in the field of thermal conductivity predictions. One example is the use of ML potentials, such as the Gaussian Approximation Potentials (GAP), in MD simulation to compute the thermal conductivity of materials like Si. Zhang et al.[138] demonstrated good agreement between the thermal conductivity values obtained from MD simulations with GAP potentials and those from DFT calculations. Another ML potential, the High Dimensional Neural Network Potential (HDNNP)[139], has shown promise in predicting the lattice thermal conductivity of semiconducting materials with the same accuracy as DFT calculations. This capability is particularly valuable when studying systems with many atoms, when DFT calculations can become computationally challenging. To overcome the limitations of inaccurate potentials in MD simulations, the development of ML potentials such as Moment Tensor Potential (MTP)[140] has been proposed. Korotaev et al.[141] utilized an ML potential called MTP coupled with MD simulations and the Boltzmann Transport Equation (BTE) method to compute the thermal conductivity of $CoSb_3$.

There have been numerous studies exploring the application of machine learning (ML) techniques in thermal transport calculations, including the solution of the phonon BTE. Li et al's work on physics-informed deep learning for solving phonon BTE with large temperature non-equilibrium is one notable example[142]. In their study, Li et al applied deep learning techniques to solve the phonon BTE in situations where there is a significant temperature difference between different regions of a material. Overall, ML techniques have the potential to accelerate materials discovery, guide the design of advanced thermal materials, and optimize energy conversion and storage devices.

## 5. COMPUTATIONAL STUDIES ON THE THERMAL TRASPORT IN 2D MATERIALS

### 5.1 Thermal Conductivity of 2D Materials: Monolayer

We briefly summarize the thermal conductivity of various 2D materials. This section mainly highlights computational findings of thermal conductivity. While some experimental results are also included. detailed experimental results are discussed in next section.

### Graphene

The monolayer graphene possesses ultra-high thermal conductivity at room temperature. The measurement of thermal conductivity depends on various factors. Experimental approaches, external conditions such as strain and substrate, sample size (length and layer thickness), and sample quality (including point defects, grain size, and impurities) all play a role in determining the measured thermal conductivity. In the study by Wei et al.[143], the thermal conductivity of suspended single-layer graphene grown by chemical vapor deposition (CVD) was reported to be approximately 2500-3100 $W/m \cdot K$ at T=350 K and 1200-1400 $W/m \cdot K$ at T = 500 K.

The contribution of different phonon modes to graphene's ultra-high thermal conductivity is a topic of debate in the nanoscale heat transfer community. The understanding of the specific phonon modes that play a crucial role in graphene's heat transfer is still a subject of ongoing research and investigation[144]. Graphene's acoustic phonon modes, including the in-plane acoustic *longitudinal wave* (LW), in-plane acoustic *transverse wave* (TW), and out-of-plane acoustic *shear wave* (SW), are indeed important in heat transfer processes. However, there are differing views regarding the significance of each mode. Some researchers argue[145] that heat transfer in graphene monolayers is primarily governed by the LW and TW modes, which are associated with the in-plane vibrations of the carbon atoms. These modes are believed to contribute significantly to the overall thermal conductivity of graphene. On the other hand, another school of thought[45, 50] emphasizes that SW mode, which involves out-of-plane vibrations of the carbon atoms, may play a more substantial role in graphene's thermal conductivity.

For practical applications, graphene will be attached to a substrate. Its in-plane thermal conductivity is often lower compared to suspended graphene. One of the reasons for this reduction in thermal conductivity is the suppression of the out-of-plane acoustic shear wave (SW) mode. For graphene monolayer on a silicon oxide substrate, the in-plane thermal conductivity of is discovered to be 600 $W/m \cdot K$ at T = 300 K[50].

There are some promising MD studies on thermal conductivity of graphene. For example, Li et al. studied thermal characteristic of graphene nanoribbons (Figure 10) endorsed by surface functionalization[146]. The thermal characteristics of graphene nanoribbons (GNRs) with surface hydrogenation were investigated using Reverse Non-Equilibrium Molecular Dynamics (RNEMD) simulations. Thermal conductivity of GNRs with fully hydrogenated domain (graphane) are studied by calculating the Kapitza conductance across the graphene-graphane interface. Thermal conductivity of hybrid nanoribbon is revealed to depend on the length as well as chirality and initial temperature and been systematically interpreted from the perspectives of morphology and phonon vibration spectra. More interestingly, remarkable thermal rectification is noticed for hybrid nanoribbon with graphene-graphane interface. Such thermal rectification decays with the length of nanoribbon and a critical length of 10 nm are identified for single layer nanoribbon beyond which the thermal rectification disappeared.

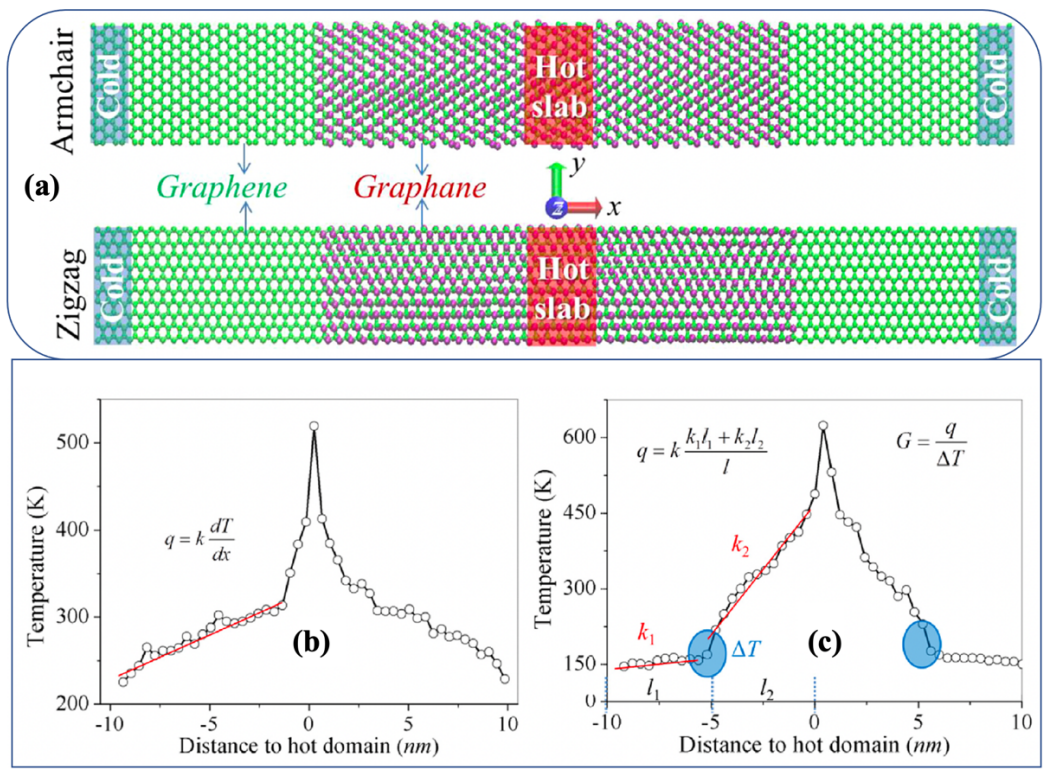

**Figure 10:** **[a]** Geometry of the RNEMD simulation for zigzag- and armchair-oriented graphene nanoribbons. The cold slabs are placed at the ends of the simulation cell, while the hot slab is in the middle of the cell. Hydrogen atoms are colored in purple while the carbon atoms are colored in green. **[b,c]** Typical temperature profile along (b) pure graphene nanoribbon, and (c) graphene nanoribbon with graphene domain under heat flux obtained by the RNEMD method. Reprinted with permission from ref[146].

Bagri et al.[147] performed nonequilibrium MD simulations to study thermal transport across twin grain boundaries in polycrystalline graphene (Figure 11). When a constant heat flux is allowed to flow, a sharp jump in temperature at the boundaries is noticed. Based on the magnitude of these jumps, the boundary conductance of twin grain boundaries was computed as a function of their misorientation angles. The boundary conductance was found to be in the range of 15-45 GW/m$^2$K, which is significantly higher than that of any other thermoelectric interfaces reported in the literature. Using the computed values of boundary conductance, they have identified a critical grain size of 0.1 $\mu m$ below which the contribution of the tilt boundaries to the conductivity becomes comparable to that of the contribution from the grains themselves.

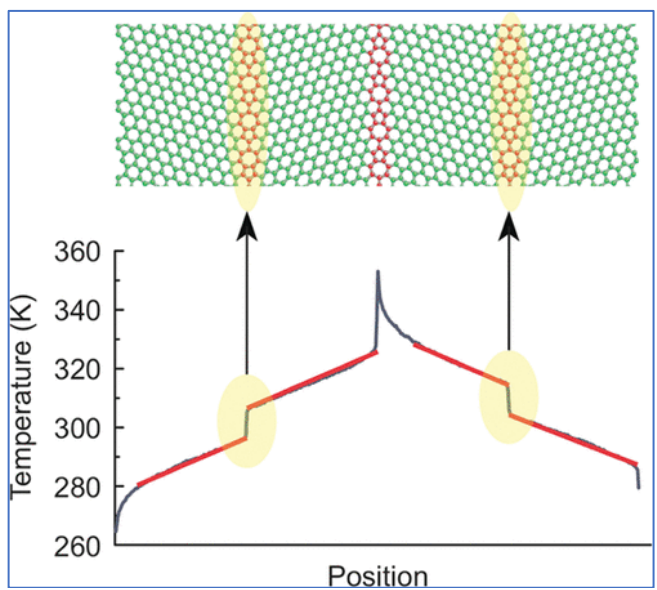

**Figure 11:** Thermal transport across Twin Grain Boundaries in Polycrystalline Graphene from Nonequilibrium Molecular Dynamics Simulations. Reprinted with permission from ref[147].

*Length Dependence:*
The intrinsic thermal conductivity ($\kappa$) of 2D systems has been predicted to exhibit a logarithmic divergence with system size[148]. However, the size dependence of $\kappa$ in 2D systems like graphene is still a topic of investigation, as the atoms can still vibrate in all three dimensions despite the material's 2D structure. The length dependence of $\kappa$ at room temperature for graphene and graphene nanoribbons (GNRs) has been explored using both NEMD and BTE methods, though the conclusions differ among studies[45, 149-152]. Convergence of $\kappa$ at a finite length was observed in NEMD simulations on supported graphene by Chen et al.[150], suggesting that $\kappa$ reaches a maximum value for a particular system size. On the other hand, Selezenev et al.[152] demonstrated a linear relationship between $1/\kappa$ and $1/L$, where $L$ represents the length of the graphene sample. Some studies have proposed a power-law relationship between $\kappa$ and length $L$ of graphene, with $\kappa \propto L^{\alpha}$, where $\alpha$ is between 0 and 1. These power-law length-dependencies have been observed in NEMD simulations[149] of graphene monolayer.

*Layers effect:*
The measurement of thermal conductivity ($\kappa$) in graphene with different layer numbers (N) provides valuable insights into the layer-dependent thermal transport properties of this material. Ghosh et al.[47] measured the $\kappa$ of graphene with layer numbers ranging from 1 to 4, as well as a multilayer graphene sample consisting of eight layers. The results of their study revealed a monotonically decreasing trend of $\kappa$ with increasing layer number.

*From Graphene to GNR: Edge Effect and Mode Quantization:*
Researchers have investigated[54, 56, 57] the chirality dependence of thermal conductivity in graphene, as thermal transport properties can vary with the specific lattice orientation or chirality. Guo et al. reported the chirality-dependent thermal conductivity in their study[149]. Studies show that:

(i) The $\kappa$ of zigzag GNRs is typically higher than the armchair GNRs. This difference in thermal conductivity arises from the distinct edge structures and phonon scattering mechanisms associated with each edge type.

(ii) The difference in $\kappa$ between zigzag and armchair GNRs diminishes as the width of GNR increases.

(iii) The $\kappa$ of GNRs tends to increase toward the bulk graphene limit as the width of GNR increases. This behavior is attributed to the convergence of thermal transport properties as the GNR width approaches the width of a bulk graphene sheet.

## Boron Nitride

The theoretical computation of single-layer boron nitride is reported[153] to be 600 $W/m \cdot K$ (by solving the Boltzmann equation).

## Molybdenum Sulfide and Other Transition Metal Sulfides

Sahoo et al.[154] reported the in-plane thermal conductivity of 11 layers of molybdenum sulfide ($MoS_2$) to be 52 $W/m \cdot K$ at room temperature. The in-plane thermal conductivity of mono- and multi-layered $MoS_2$ at room temperature[155] is 35-52 $W/m \cdot K$. However, some other researchers predict in-plane thermal conductivity of mono- and multi-layered $MoS_2$[156]. The reported result is 77-84 $W/m \cdot K$, which is much larger than prior reported data. Therefore, there are many ambiguities in thermal conductivity measurement of $MoS_2$.

The lower thermal conductivity of monolayer $MoS_2$ compared to graphene and boron nitride can be attributed to several factors. *(i)* Theoretical calculations reveals that monolayer $MoS_2$ has lower phonon group velocity and a larger Gruneisen constant, which are primary factors contributing to its low thermal conductivity. These properties result in a shorter mean free path of phonons in $MoS_2$ (merely 14.6 nm), limiting their ability to transport heat over long distances. *(ii)* In monolayer $MoS_2$, the contribution of out-of-plane acoustic *shear wave* (SW) to thermal conductivity is lower than the in-plan acoustic modes. In contrast, in planar 2D materials like graphene and boron nitride, the contribution of the SW mode to thermal conductivity can be more significant, often accounting for more than 50% of the total thermal conductivity.

## Black Phosphorus, Black Arsenic, Phosphorene

Black Phosphorous (BP) is a promising 2D material with a unique "Great Wall" structure that introduces in-plane anisotropy, which can result in anisotropic thermal conductivity[157-160]. The "Great Wall" structure of BP consists of vertically stacked layers held together by van der Waals forces, leading to pronounced anisotropic properties. The out-of-plane phonon mode has a positive impact on thermal conductivity because of "Great Wall" structure. Theoretical studies have reported an internal thermal conductivity ratio of approximately 30.15 $W/m \cdot K$ in zigzag (ZZ) direction and 13.65 $W/m \cdot K$ in armchair (AC) direction[161]. Other theoretical calculations predicted similar values[160].

**Telluride**

Two-dimensional telluride materials have emerged as promising candidates for thermoelectric applications[162, 163]. First principles calculations have provided insights into the thermal conductivity and thermoelectric properties of monolayer telluride with various structures[164,165]. Experimental studies with Raman method measured thermal conductivity of telluride in intra-chain direction is only[166] $1.5\ W/m \cdot K$.

**Silicene**

Silicene, a 2D material composed of a single layer of silicon atoms, was first synthesized in 2012. However, the growth condition and stability conditions of silicene are challenging[167, 168]. Therefore, not much progress happened on the research on thermal conductivity measurement of silicene.

**Other 2D Materials**

There has been significant interest in exploring the thermal conductivity of various novel 2D materials. Theoretical studies on thermal conductivity measurement have been performed on various 2D materials such as selenium oxide[169], gallium nitride[170], tin sulfide[171], zirconium telluride[172], bismuth telluride[173], indium selenide[174], etc.

## 5.2 Thermal Conductivity of 2D Materials: Homo- and Hetero-structures

The fundamental understanding of thermal transport in 2D heterostructures is crucial for the development of next-generation energy storage and other applications. Heterostructures are divided into two classes:
(i) *In-plane heterostructures* involves the seamless stitching of two 2D materials into a shared plane, creating an atomically sharp 1D interface.
(ii) *van der Waals (vdW) heterostructures* consist of vertically stacked layers of distinct 2D materials, with non-bonded vdW forces acting between adjacent layers.

Thermal conductivity of monolayer 2D materials can differ from that of their corresponding heterostructures. For instance, thermal conductivity of graphene and graphene nanoribbons (GNRs) generally decreases with tensile strain due to phonon softening[175-177]. In contrast, the tensile strain in graphene/h-BN in-plane heterostructures can significantly enhance the interfacial thermal conductance by improving the alignment of out-of-plane phonon modes near the interface[178]. Furthermore, 2D heterostructures exhibit various abnormal thermal phenomena, sparking debates and research in the field. These phenomena include Thermal Rectification (TR)[179, 180], Negative Differential Thermal Resistance (NDTR)[181], Thermal Bandgap[182], Thermophoresis[183]. It is worth noting that while graphene has been extensively studied and has a high thermal conductivity, the knowledge and understanding of thermal transport properties in graphene may not be directly applicable to other 2D materials. Each material has its own crystal structure, atomic species, and phonon dispersion characteristics, leading to different thermal transport behaviors.

### 5.2.1 In-Plane Heterostructures

In-plane heterostructures require matched lattice structures of the constituent materials. The presence of interfaces can introduce novel properties that differ from the parent materials. Measuring the in-plane thermal conductivity (ITC) of in-plane heterostructures is indeed challenging, especially compared to vdW interfaces. Therefore, theoretical simulations are urgently necessary. Hong et al.[184] studied the effect the sample length and ambient temperature on the ITC of graphene/h-BN in-plane heterostructures *via* NEMD method. The results showed that the ITC varies from 1.9 to 4.5 GW/m$^2$K as the sample length ranged from 20 to 100 nm. Additionally, it was observed that increasing the temperature significantly enhanced the ITC. However, in-plane heterostructures are not significantly used for energy storage. Therefore, we will not further explore this scenario.

**5.2.2 van der Waals (vdW) Heterostructures**

The vdW heterostructures provide a unique advantage in creating ultra-clean interfaces without the need for strict lattice-matching constraints. These heterostructures offer novel ways to manipulate the properties of emergent devices through factors such as the stacking sequence, relative rotation between adjacent layers, and interlayer spacing. The thermal properties of vdW heterostructures are of great interest as they play a crucial role in determining the functionality, performance, and reliability of such heterostructures-enabled devices. The phonon dispersion analysis[185] has shown that the AB stacking configuration is superior to the AA stacking for graphene/h-BN vdW heterostructures in terms of thermal properties. NEMD calculations[36] have indicated that the thermal conductivity of h-BN-supported graphene is only 23% lower compared to the suspended case, suggesting that the h-BN substrate has minimal impact on the force environment of graphene. Additionally, researchers have observed that the thermal transport properties of vdW heterostructures strongly depend on the stacking configurations. Wu et al. studied the interfacial thermal conductance across graphene/MoS$_2$ vdW heterostructures by first principles and molecular dynamics (MD) simulations. The predicted thermal conductivity (k) of monolayer graphene and monolayer MoS$_2$ reaches 1458.7 W/mK and 55.27 W/mK, respectively. Furthermore, the thermal conductance across the graphene/MoS$_2$ interface was calculated to be 8.95 MW/m$^2$K at 300 K. This value represents the ability of heat to transfer across the interface between the two materials.

The thermal conductivity and interfacial thermal conductance (ITC) of graphene-based vdW heterostructures can be significantly influenced by various factors, including temperature, hydrogenation, and interlayer coupling. At low temperatures, the thermal conductance of graphene coupled with h-BN decreases, but it can still retain up to 96% of its value at room temperature[186]. As the temperature increases, the in-plane thermal conductivity of the heterostructures decreases, while the ITC increases. Hydrogenation of graphene in graphene/h-BN vdW heterostructures has been observed to improve the ITC by more than 70%[187]. The hydrogenation enhances the coupling between in-plane and out-of-plane phonons and widens the phonon channels, thereby enhancing the thermal transport across the interface. In case of graphene/phosphorene vdW heterostructures, constructing a sandwiched structure with graphene layers can enhance the thermal stability and thermal conductivity of phosphorene[188]. The in-plane thermal transport in these heterostructures exhibits anisotropy and increasing the interlayer coupling can enhance the thermal conductivity of the phosphorene layer due to phonon stiffening[115]. The ITC of graphene/phosphorene/graphene multilayer structures can be influenced by defect, hydrogenation, compressive strain, as observed in NEMD simulations[189].

The room temperature thermal conductivity of $MoS_2/MoSe_2$ bilayer heterostructure was predicted[190] to be 25.4 W/mK using the BTE method. In the graphene/$MoS_2$ bilayer heterostructure, the interlayer coupling can decrease the in-plane thermal conductivity of graphene, while it has a minimal effect on the thermal transport in the $MoS_2$ layer[191]. The ITC in heterostructures can be modulated by various factors, including temperature, contact pressure, hydrogenation, and vacancy defects. In the case of $MoS_2$/h-BN vdW heterostructure, the observed ITC (~ 17 MW/m$^2$K) is approximately one-third of the ITC observed in graphene/h-BN heterostructures (~ 52 MW/m2K). Why does graphene/h-BN have more ITC? The lighter mass of carbon atoms in graphene compared to other elements, such as boron and nitrogen in h-BN, leads to a wider range of phonon frequencies available for transmission between the graphene and h-BN regions.

**5.2.3 Modification of Thermal Transport in 2D materials and their heterostructures**

Controlling and manipulating phonon transport in 2D heterostructures is indeed a complex task due to the diverse interactions and broad frequency spectrum of phonons involved. However, by introducing structural defects such as isotopes, vacancies, dislocations, and utilizing substrates chemisorption, and strain, the phonon scattering processes and thermal transport can be tuned and controlled. In the development of 2D heterostructures for energy storage, heat dissipation is a critical consideration. The generation of waste Joule heat can lead to the formation of high-temperature hot spots in devices, which can significantly affect their stability and performance. To mitigate this issue, designing heterostructure arrangements that effectively dissipate heat and prevent thermal runway is crucial. The choice of 2D building blocks with high thermal conductivity can aid in alleviating Joule heat dissipation and improving heat dissipation ability in energy storage.

*Doping*
During the preparation of 2D materials, imperfections such as isotopes, point-defect, and grain boundaries are often unavoidable and can have a significant impact on the thermal conductivity of material. These imperfections act as scattering centers for phonons, leading to a decrease in thermal conductivity. For example, in GNRs with SW defects, interesting phonon transport behaviors have been observed[192]. One such behavior is the presence of edge heat flux, where heat is preferentially conducted along the edges of the GNRs. Additionally, the appearance of circulating heat flux near SW defects has also been observed, indicating complex phonon scattering processes in the presence of defects. In case of Si-doped graphene[193], the presence of Si dopants significantly reduces the probability of phonon transmission across the doped region. This results in a significant reduction in thermal conductivity due to the increased scattering of phonons by the dopant atoms.

*Defect*
NEMD simulations[194] revealed that thermal conductivity of $MoS_2$ is reduced by over 60% for a 0.5% mono-Mo vacancy concentration.

*2D alloy*
In the case of 2D alloys, combining different 2D materials or introducing dopant atoms can create new materials with altered thermal properties[195]. The presence of different elements or dopants can introduce additional phonon scattering centers, disrupting the phonon propagation, and altering the thermal conductivity.

*Defect at the interface*
The introduction of vacancy defect at the interface layers can indeed enhance ITC[196]. Vacancy defects can create additional scattering centers for phonons, leading to increased phonon coupling and enhanced thermal transport at the interfaces. By using the thermal bridge method, Aiyiti et al.[197] demonstrated that defect is an effective approach to lower the thermal conductivity of few layer $MoS_2$.

*Mechanical strain vs thermal conductivity*
The mechanical strain is a powerful tool for tuning the properties of 2D materials, including thermal conductivity. In general, the thermal conductivities of 2D planar nanostructures (*e.g.,* graphene, graphyne) tends to decrease with increasing tensile strain. This is because strain can induce phonon softening, which leads to a decrease in phonon group velocity[176] and a reduction in thermal conductivity. Similarly, for materials like $MoS_2$, an increase in tensile strain usually results in a decrease in thermal conductivity[198]. However, the relationship between strain and thermal conductivity can also depend on the sample length. Studies[199] have shown that for graphene, the applied tensile strain can either reduce or increase the thermal conductivity, depending on the sample length. The interplay between strain-induced phonon scattering and phonon transport across the sample length can lead to different strain-dependent behaviors. For the buckled 2D materials like silicene or phosphorene, the dependence of thermal conductivity on tensile strain can exhibit a transition from an initial increase to subsequent reduction variation. This behavior is attributed to the interplay between strain-induced phonon softening and changes in the phonon dispersion relations.

*Strain on heterostructure*
For graphene/h-BN superlattice, using NEMD, Zhu et al.[200] discovered that the thermal conductivity initially increases with strain, reaching to a peak value, and then decreases with further strain. This behavior can be attributed to the interplay between strain-induced phonon scattering and changes in the interlayer coupling. In case of graphene/black phosphorous vdW heterostructures, compressive strain of 0.18 increases ITC more than 15-fold compared to the strain-free value[201]. The applied compressive strain strengthens the interlayer coupling, improving the phonon transport between layers. Additionally, the enhanced shear interaction between interlayers further contributes to the thermal transport from in-plane phonons. This leads to a substantial increase in ITC compared to the strain-free case. Guo et al. studied tensile strain dependence of the thermal conductivity of monolayer graphene. They found that under 0.12% biaxial tensile strain, the system's thermal conductivity drops approximately by 20% at 350 K, which verifies the previous theoretical prediction[202].

*Chemical functionalization*
Chemical functionalization offers an effective approach to modulate the thermal conductivity of 2D heterostructures. The use of hydrogenation has been shown to have a significant impact on phonon transport. BTE calculations[203] have revealed that hydrogenation can enhance the phonon transport in pentagraphene, leading to a substantial increase in thermal conductivity, up to 76% compared to the non-hydrogenated state. In case of graphene/phosphorene bilayer heterostructure, the interfacial thermal resistance (ITR) increases monotonically with the hydrogen coverage in graphene layer. When 2D materials or related heterostructures are employed in device applications, their thermal transport properties are influenced by the phonon-substrate scattering. The heat generated is dissipated directly into the substrate through their interfaces. The presence of the substrate can lead to enhanced optical phonon scattering in the 2D material[204], thereby affecting its overall thermal conductivity.

# 6 EXPERIMENTAL DETAILED STUDIES ON THERMAL TRASPORT IN 2D MATERIALS

The experimental measurement of thermal transports, including thermal conductivity, diffusivity, and related properties, plays a crucial role in advancing our knowledge of heat transfer mechanisms, optimizing energy conversion and storage systems, and developing innovative thermal management strategies. Accurate experimental measurements enable researchers to validate theoretical models, identify the underlying physical mechanisms governing thermal transport, and guide the design of novel materials and devices with tailored thermal properties. Moreover, research in this field holds immense significance in addressing critical challenges, including thermal management in electronics, thermoelectric energy conversion, thermal insulation, and heat dissipation in nanoscale systems.

In 2010, Cai et al.[143] conducted a study (Figure 12) focused on measuring the thermal conductivity and thermal interface conductance of monolayer graphene using a micro-Raman spectroscopy approach. They utilized a large-area, high-quality monolayer graphene grown via chemical vapor deposition (CVD) on 25 µm thick Cu foils as their sample material. To facilitate thermal measurements, a CVD-grown graphene was placed on an Au-coated SiNx membrane featuring a 100 × 100 array of 3.8 µm diameter holes with a 10 µm pitch between holes. The quality of the graphene was verified through micro-Raman spectroscopy, which confirmed the presence of suspended graphene on the holes by Raman mapping image of the G peak intensity and the absence of the D band associated with defects. To perform the measurements, a 532 nm laser beam was directed at either the center of the suspended graphene flake or the supported graphene area on the Au/SiNx membrane using an objective lens. The transmission of the laser beam through the suspended graphene was measured using a semiconductor laser power meter positioned under the SiNx support, as depicted in Figure 12.

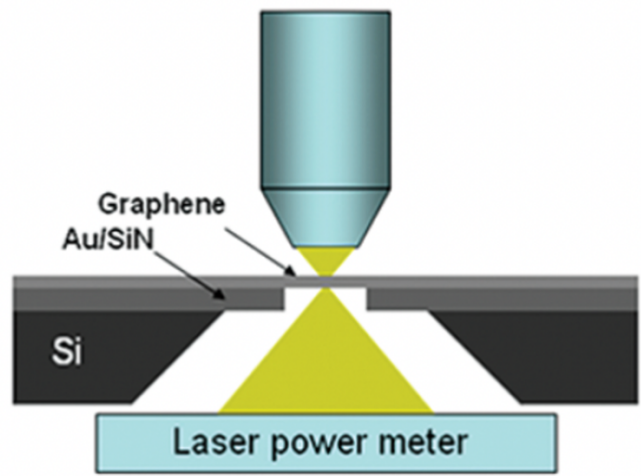

**Figure 12:** Schematic of the micro-Raman spectroscopy based experimental setup for the measurement of thermal conductivity of graphene by Cai et al. [9]

The power absorbed by the suspended graphene was determined by comparing the power transmitted through an empty hole to that transmitted through the graphene flake. By considering the red shift of the Raman G peak in response to temperature rise due to bond softening, and the linear dependence between the red shift and sample temperature, Cai et al. calibrated the relationship between the Raman shift and temperature rise. They achieved this by obtaining Raman spectra of the graphene sample while it was placed on a heating stage with its temperature measured by a thermocouple. The resulting data revealed a downshift of the Raman G peak at a rate of $(4.05 \pm 0.2) \times 10^{-2}$ cm-1/K. Utilizing this calibration, the researchers plotted the downshift of the Raman G peak and temperature as a function of absorbed laser power, as shown in Figure 13. Through subsequent analysis, the temperature distribution of the suspended graphene under laser heating was determined, allowing for the generation of a plot illustrating the thermal conductivity of suspended graphene as a function of temperature. This plot, presented in Figure 14, also included a comparison of the obtained data with thermal conductivity values for various graphite samples reported in the literature.

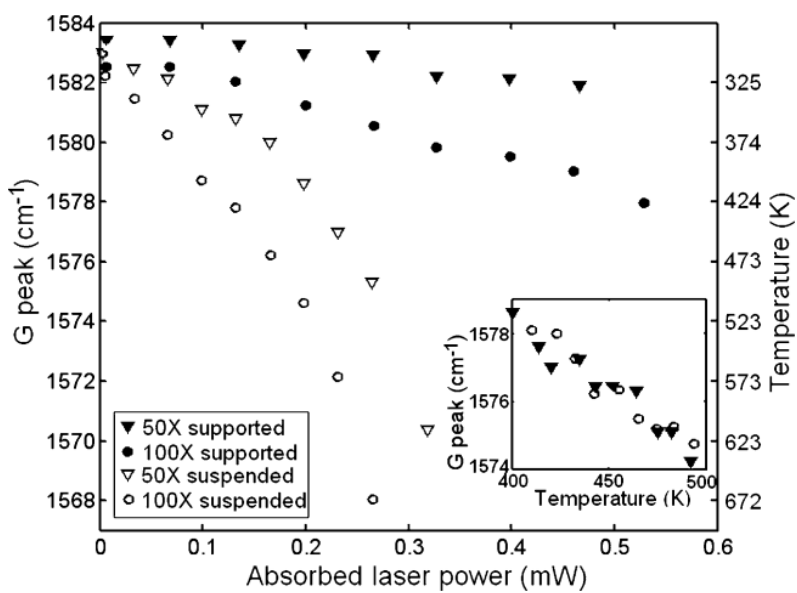

**Figure 13:** The G peak shift (left axis) and temperature (right axis) measured on the supported graphene and at the center of the suspended graphene as a function of the absorbed laser power; by Cai et al.

Wang et al. employed a micro electrothermal system to experimentally measure the thermal conductivity of both suspended and few-layers of supported graphene[205]. Building upon the techniques developed by Li et al.[206] and Shi et al.[39], they utilized microfabricated electrothermal systems, as depicted in Figure 15 (a and b), to facilitate their measurements. The graphene layers were carefully deposited onto the measurement system, bridging the gap between the heater and sensor components. The deposition of the graphene layers onto the system was achieved using the electron beam lithographic technique. To conduct the thermal measurements, the graphene layer was heated by passing a direct current through the heater loop. By measuring the electrical resistances of both the heater and sensor loops and calibrating them with temperature variations, the temperatures of the loops were obtained which allowed them to calculate the heat conductance of the graphene layer. Subsequent analysis was then carried out to investigate the thermal conductivity of the graphene layers as a function of temperature.

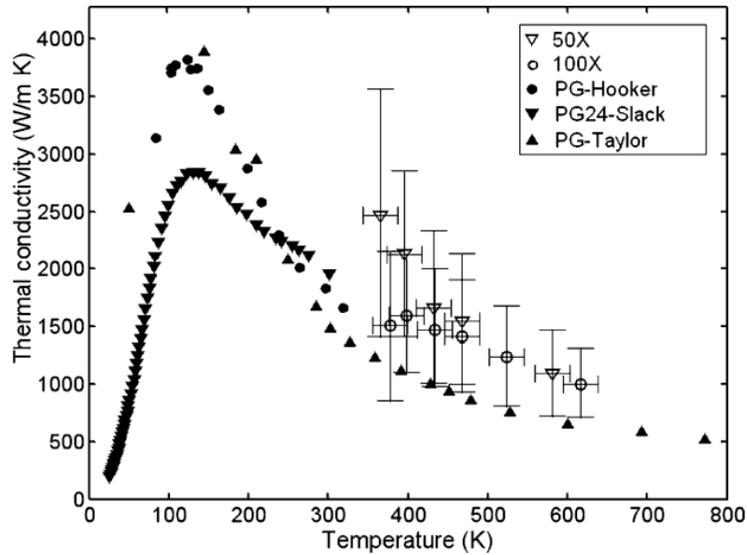

**Figure 14:** Thermal conductivity of the suspended CVD graphene measured using the 100× and 50× objective lens as a function of the measured graphene temperature; measured by Cai et al.[9]. Also shown in comparison are the literature thermal conductivity data of pyrolytic graphite samples.

Hexagonal boron nitride (h-BN) is a promising two-dimensional material with intriguing thermal transport properties. Its unique lattice structure composed of alternating boron and nitrogen atoms gives rise to exceptional thermal conductivity, even surpassing that of graphene in specific occasions. Experimental studies have revealed that h-BN exhibits an anisotropic thermal conductivity, with significantly higher values in the in-plane direction compared to the out-of-plane direction. This anisotropy is attributed to the strong covalent bonding within the h-BN layers and weaker van der Waals interactions between the layers. The excellent thermal properties of h-BN make it a desirable candidate for applications in thermal management, heat dissipation, and nanoelectronics, where efficient heat transfer is crucial.

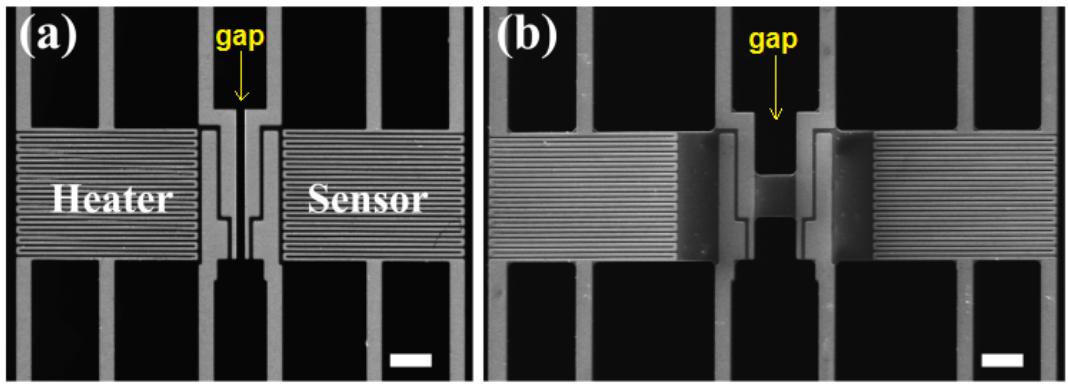

**Figure 15:** SEM images of the micro electrothermal system for suspended and supported graphene samples; scale bar: 5 μm.

Jo et al. conducted an experimental study to measure the thermal conductivity of suspended few-layer hexagonal boron nitride (h-BN) using a microbridge device equipped with resistance thermometers[207]. They adapted a previous design by Seol et al.[50] intended for assessing 2-D phonon transport in supported graphene. To create the microbridge device with suspended h-BN layers, they employed a multistep process: (1) transferring and suspending few-layer h-BN samples onto a central frame made of SiNx beams; (2) fabricating a rectangular h-BN ribbon through electron-beam lithography (EBL) and plasma etching, with Au registration marks for reference; (3) attaching the suspended h-BN sample and gold alignment marks to a PMMA film, aligning it with the microbridge device, and annealing for enhanced adhesion. The PMMA film was then dissolved, resulting in a fully assembled device featuring an 11-layer suspended h-BN sample. By analyzing the electrical resistances at various points of the system and their relationship with temperature changes induced by electrical heating, the thermal conductivity of the suspended few-layer h-BN samples were determined.

Experimental measurements of thermal transport properties in various 2D materials are commonly conducted using microfabricated devices. These devices generally consist of two parts: a heater loop and a sensor loop, separated by a micro gap. Platinum or platinum-based alloys are typically patterned onto a suspended silicon nitride (SiNx) membrane, which is a few hundred nanometers thick, to create these loops. The desired 2D materials are then positioned on the device surface, forming a bridge between the loops; electron beam lithography is a commonly employed technique for this purpose. The 2D material is heated by passing an electric current through the heater loops, causing heat to be conducted through the material and reach the sensor loop. Temperatures at different points are measured following an appropriate equivalent thermal circuit. By measuring the electrical resistances of the loops and their calibrated relationship with temperature variations, the thermal conductivity of the 2D material under investigation can be obtained as a function of temperature. Another heating approach observed in similar microdevices in the literature involves directly applying an electron beam to the suspended 2D material. Micro-Raman spectroscopy is often employed to experimentally measure the thermal transport properties of 2D materials. By subjecting the 2D material to a laser beam, an observable redshift is observed in its Raman spectra as the temperature of the material increases. Through a calibrated relationship between the red shift in the Raman spectra, laser power, and temperature, it becomes possible to measure the thermal transport properties of the 2xD material.

Lastly, we summarize some more experimental observation of thermal conductivity and interfacial thermal conductance (ITC) for different materials.

1. Graphene Monolayer: Baladin et al. measured thermal conductivity of graphene monolayer using the optothermal Raman technique, obtaining values ranging from 2000 to 5000 $W/m \cdot K$ at room temperature[37].
2. Bilayer Graphene: The thermal conductivity of bilayer graphene was also measured[208] and compared with the theoretical predictions[209].
3. Defective Graphene: Malekpour et al.[210] reported that the room-temperature thermal conductivity of defective graphene decreases from 1800 to 400 $W/m \cdot K$ as the defect density induced by electron beam irradiation slightly increases from 0.0005 to 0.005%.
4. Suspended Graphene: Xu et al.[211] discovered a logarithmic dependence of thermal conductivity on the sample length ($\kappa \sim \log L$) in suspended graphene using the microbridge technique.

5. Few-Layer Graphene: There is positive correlation between thermal conductivity and number of layers in few-layer graphene[212] and the value of ten-layer graphene is over 1000 $W/m \cdot K$.
6. $MoS_2$: Jo et al. performed optothermal Raman measurement and determined thermal conductivity of 4-layer $MoS_2$ is 44-50 $W/m \cdot K$ at room temperature[207].
7. Interfacial Thermal Conductance (ITC): The ITC measurement in in-plane heterostructures is challenging compared to that of vdW interfaces. The measured ITC of graphene/substrate interface is reported to be in the range of 20 - 100 MW/m$^2$K at 300 K. Lee et al.[171] utilized TDTR-based approach and reported ITC of $MoS_2$/metal thin-films interface is ~ 26 MW/m$^2$K.

## 7 POTENTIAL PROBLEMS ON THERMAL TRASPORT IN 2D MATERIALS-BASED ELECTROCHEMICAL ENERGY STORAGE SYSTEMS

*How thermal conductivity of 2D materials vary with intercalation and deintercalation?*

Figure 6 shows that PGN can adsorb more intercalating ions. DFT studies compliment the experimental findings. The thermal conductivity of 2D materials can indeed vary with different stages of ion adsorption/desorption. The presence of intercalating ions can affect the thermal transport properties of the material due to changes in the lattice dynamics and phonon scattering mechanisms.

During ion adsorption, the intercalating ions can introduce additional mass and alter the atomic vibrations within the lattice. This can lead to increased phonon-phonon scattering, which hinders the propagation of heat-carrying vibrations (phonons) and consequently reduces the thermal conductivity of metals. The specific effect on thermal conductivity depends on factors such as the type of ions, their concentration, and the interaction strength between the ions and the 2D material. Conversely, during ion desorption, the removal of intercalating ions can restore the original lattice structure and phonon dynamics, potentially resulting in the recovery or even enhancement of the thermal conductivity.

Furthermore, the distribution of ions over the 2D material, including graphene, can significantly impact thermal transport. The arrangement and density of ions affect the phonon scattering pathway and the overall thermal resistance of the system. Different ion distribution can lead to distinct phonon scattering mechanisms, resulting in variations in thermal conductivity. Defects in the 2D material's structure also play a crucial role in thermal transport. Defects can act as scattering centers for phonons, impeding their propagation and reducing thermal conductivity. The combination of different defect topologies with varying ion distributions further complicates the analysis of thermal transport in 2D materials.

While graphene has been extensively studied, it is essential to conduct similar analyses for various other 2D materials, such as transition metal dichalcogenides (TMDs) and MXenes, as their properties may differ. The interplay between ion adsorption, defect structures, and material-specific characteristics will determine the thermal conductivity behavior during different stages of ion adsorption/desorption in these materials. Theoretical studies, along with experimental investigations, can provide valuable insights into understanding and optimizing thermal transport in diverse 2D materials.

*Thermal conductivity of 2D materials used as vdW slippery interface*

Figures 7,8 demonstrate 2D materials graphene as vdW slippery surface over the current collector. This arrangement reduces the interfacial stress. However, proper thermal transport between current collector and electrode is important for efficient heat dissipation. The specific changes in thermal transport characteristics depend on factors such as the thickness and quality of the graphene layer, the interface properties between graphene and the current collector, and the nature of phonon coupling across the interfaces.

When considering different material combinations, such as graphene/copper, graphene/nickel, $MoS_2$/copper, $MoSe_2$/copper, etc., the thermal transport behavior will vary based on the thermal conductivities and interfacial properties of the individual materials involved. However, the study of thermal transport in systems involving electrodes, 2D materials, and current collectors is indeed challenging. Computational approaches, such as molecular dynamics simulations and density functional theory calculations, can be employed to investigate thermal transport in these systems. These studies need to consider factors such as intercalation/deintercalation of electrodes, which can affect the phonon scattering mechanisms and the overall thermal conductivity of the system. Additionally, the presence of defects, both in the 2D materials and at the interfaces can significantly influence thermal transport. Defect can act as scattering centers for phonons. Impending their propagation and reducing the overall thermal conductivity.

In summary, the thermal transport behavior changes when conventional current collectors are covered with 2D materials such as graphene. The specific changes depend on the material combinations, interfacial properties, defect topologies, and the presence of intercalating/deintercalating electrodes. Computational methods play a key role in studying and understanding the complex thermal transport characteristics of these systems.

*Thermal transport of 2D materials as conductive additives*

Figure 5b2 shows the integration of 2D materials and active materials in various ways – (i) *encapsulated*, (ii) *anchored*, (iii) *wrapped*, (iv) *layered*, (v) *mixed*, (vi) *sandwich-like*. Proper heat dissipation through these 2D materials is necessary for thermal runway prevention. Moreover, 2D materials need to be in different curvatures, topologies for efficient design of conductive additives. The thermal conductivity of 2D materials can vary for different curvatures when used as additives. Curvature can affect the phonon scattering mechanism and the overall phonon transport in 2D materials, thereby impacting their thermal conductivity.

The contact of additives with 2D materials can also influence their thermal properties. When 2D materials are in contact with other materials, such as active materials or electrodes, the interfacial thermal resistance plays a crucial role in determining the overall thermal transport behavior. The efficiency of heat transfer across the interface depends on factors such as the nature of the contact (e.g., direct contact or via a bonding layer), the interfacial properties (e.g., bonding strength, lattice matching) and the presence of any interfacial defects or impurities. The experimental investigations and computational techniques can provide valuable insights into understanding the thermal conductivity variations of 2D materials with different curvatures and the effects of interfacial contact on their thermal properties.

*Thermal transport of 2D materials used for controlling electrode-electrolyte interface*

The influence of electrolyte contact on the thermal conductivity of 2D materials is a complex topic that has not been extensively studied. However, there are some general considerations and potential effects to be aware of:

1. *Intermolecular Interactions*: When 2D materials are in contact with electrolytes, intermolecular interactions can occur between the electrolyte molecules and the atoms or functional groups on the surface of 2D material. These interactions can affect the phonon scattering processes and alter the thermal transport properties. Depending on the strength and nature of these interactions, the thermal conductivity of the 2D material may be influenced.

2. *Ion Interactions*: Electrolytes consist of ions in solution, and these ions can interact with the 2D material's surface. The presence of ions near the surface of the 2D material can affect the phonon scattering mechanisms, introducing additional scattering centers, and altering the thermal conductivity. The specific impact of ion interactions on thermal transport depends on factors such as ion size, charge, and concentration.

3. *Electrolyte Composition:* Electrolytes can be organic or aqueous, and different electrolyte compositions can have distinct effects on thermal transport. Organic electrolytes often contain solvents such as dimethyl carbonate (DMC) or ethylene carbonate (EC), while aqueous electrolytes can include salts like lithium bis(trifluoromethanesulfonyl)imide (LiTFSI). The specific chemical properties of these electrolytes, such as their molecular structure, viscosity, ionic conductivity, and thermal conductivity, can potentially influence the thermal transport properties of 2D materials.

The impact of electrolytes on thermal transport in 2D materials is still an area of ongoing research, and more comprehensive studies are needed to understand the specific effects. Experimental investigations, along with computational simulations, can provide insights into the interplay between electrolytes and thermal conductivity of 2D materials.

*Thermal transport of 2D materials used for controlling cathode-electrolyte interface*

Traditional cathode materials are Lithium Cobalt Oxide (LiCoO$_2$), Nickel Cobalt Aluminium Oxides (NCA). 2D materials are also used for encapsulation of cathode materials. In this case, it is important to study thermal transport variation of 2D materials in contact with cathode materials. In addition, how will thermal conductivity of 2D materials varies when in contact with electrolytes?

*Thermal transport of 2D materials used for separators*

2D materials-coated separators or 2D materials as separator should prevent electrical contact between the cathode and anode without compromising the ion diffusion. However, separator need to dissipate heat as well to prevent thermal runway. To maintain diffusivity, most 2D materials may need to have porous structures, e.g., vacancy defect in graphene.

*Thermal transport of 2D heterostructure*

For any heterostructures for potential energy storage applications, first and foremost, it is necessary to understand whether the structure is stable or not. Tungsten disulfide ($WS_2$) and diselenide ($WSe_2$) heterostructures ($WS_2$- $WSe_2$) are considered in Figure 16[213]. Figure 16a and d show the *chalcogen* (*c-*) and *metal* (*m-*) terminated $WS_2$-$WSe_2$. For each termination, AA- (Figure 16b1, e1) and AB-stacking (Figure 16c1, f1) systems are considered.

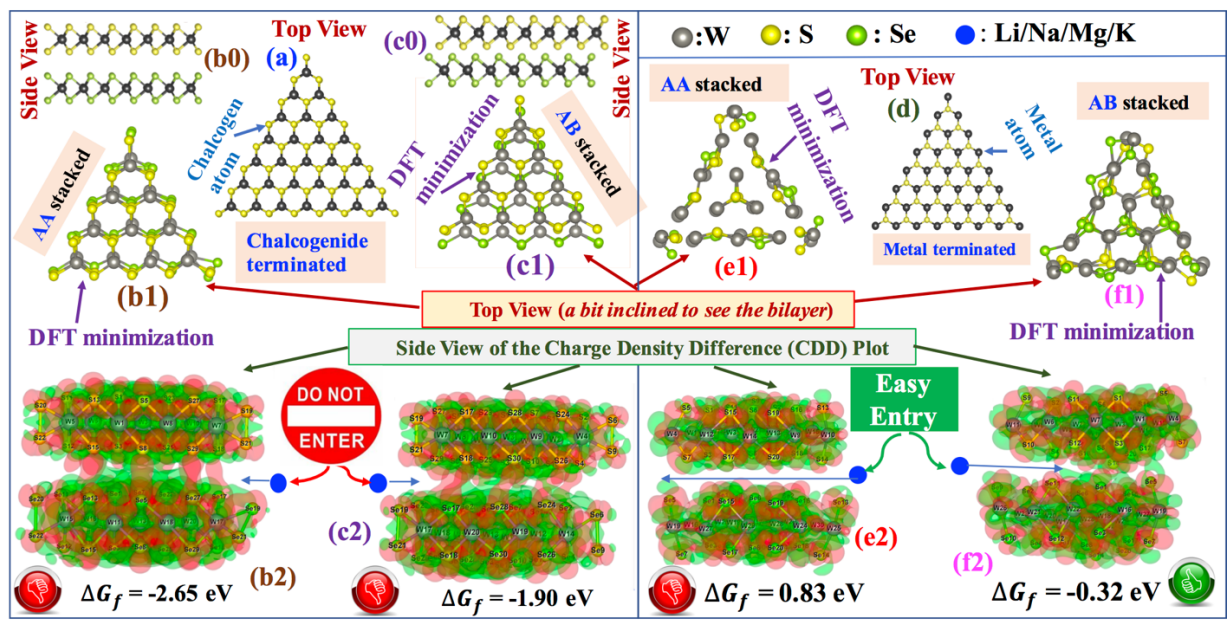

**Figure 16:** DFT study of stability of $WS_2$/$WSe_2$ heterostructures considering various edge terminations and stacking for energy storage

DFT minimization, Gibbs free energy ($\Delta G_f$), and charge-transfer analysis were performed for all four cases. From the DFT minimized structure, we conclude that *c*-terminated AA-stacked (Figure 16b1) is the most stable structure ($\Delta G_f$ = -2.65 eV) while *m*-terminated AB-stacking is the least stable ($\Delta G_f$ = -0.32 eV). AA-stacking for *m*-termination is not stable since $\Delta G_f$ is positive (($\Delta G_f$ = 0.83 eV). Charge Density Difference (CDD) analysis shows that for the most stable structure (Figure 16b2), charge overlaps between layers indicating strong interlayer interaction. Charge overlap reduces as the stability reduces. ***For the <u>energy storage purpose</u>***, fast ion transport and kinetics between the layers of 2D materials is necessary. Charge-overlap acts as "*Do Not Enter*" zone blocking the fast transport (Figure 16b2, c2). No charge overlap corresponds to an unstable structure (Figure 16e2), which can't be used for energy storage. Hence, intermediate stable structure with less charge-transfer is most preferable (Figure 16f2). The concept is analogous to $W_{sep}$ between bulk and 2D materials (Figure 7), where very high or low $W_{sep}$ is not good for energy storage application.

The above analyses showcase the configuration suitable for ion intercalation. However, this configuration may not be the best one for thermal transport/heat dissipation. Therefore, thermal analyses are necessary for various heterostructures combinations considering different surface terminations, stacking, etc. The analyses become more complicated considering the presence/effect of intercalation ions, electrolytes. The previous sections describe some open problems in 2D materials based EESS. However, there are many computational challenges involved in achieving these goals. Molecular Dynamics simulation needs reliable accurate interatomic potentials. However, suitable potentials are unavailable for many systems. For example, consider a problem of studying thermal conductivity of graphene in contact with aqueous electrolyte LiTFSI. For MD simulation, suitable potential is needed to model graphene/LiTFSI, which is currently unavailable. Density Functional Theory (DFT) calculation is computationally very expensive. Machine learning approach is urgently necessary for faster prediction of thermal conductivity of low-dimensional materials-based EESS.

## 8 CONCLUSIONS

We have addressed the importance of studying the thermal conductivity of 2D materials and their heterostructures for energy storage and conversion applications. The perspective review summarized several pioneering studies of thermal transport in several 2D materials, including factors such as topology, external loading, doping, etc. However, none of these studies are in the context of energy storage. The energy storage community can import some existing results on 2D materials' thermal transport properties for designing 2D materials-based energy storage devices. However, we highlighted that during practical applications, several factors, such as ion intercalation/deintercalation and electrolyte effect, may drastically alter the thermal transport behaviors of 2D materials. Therefore, studying the thermal transport of 2D materials in the context of heat transfer is essential. There are many computational and experimental challenges involved, given the complexity of the problems. Therefore, computational communities from several sectors, such as Density Functional Theory (DFT), Molecular Dynamics (MD), and Machine Learning, need to work on these problems. Moreover, synergistic experimental-computational collaborative efforts are essential for realistic outcomes for the practical design of 2D materials-based energy storage devices. The perspective review is expected to spark new research directions in the heat transfer community.

**AUTHOR INFORMATION**

**Corresponding Author**

Dibakar Datta

Email: dibakar.datta@njit.edu, Phone: +1 973 596 3647

Eon Soo Lee
Email: eonsoo.lee@njit.edu, Phone: +1 973 596 3318

**Author Contributions**
D.D. and E.S.L. conceived the project and wrote the paper. All authors approved the final version of the manuscript.

**CONFLICT OF INTEREST STATEMENT**

The authors have no conflicts of interest to declare. All authors have seen and agree with the contents of the manuscript and there is no financial interest to report. We certify that the submission is original work and is not under review at any other publication.
**ACKNOWLEDGEMENT**

The work is supported by National Science Foundation (NSF) (Award Number # 2126180 and #2237990). Authors acknowledge Advanced Cyberinfrastructure Coordination Ecosystem: Service & Support (ACCESS) for the computational facilities (Award Number - DMR180013).